\documentclass[english,aps, twocolumn]{revtex4-1}
\usepackage[T1]{fontenc}
\usepackage[latin9]{inputenc}
\usepackage{color}
\usepackage{babel}
\usepackage{amsmath}
\usepackage{amssymb}
\usepackage{graphicx}
\usepackage[unicode=true,pdfusetitle,
 bookmarks=true,bookmarksnumbered=false,bookmarksopen=false,
 breaklinks=false,pdfborder={0 0 0},pdfborderstyle={},backref=false,colorlinks=true]
 {hyperref}
\hypersetup{
 linkcolor=blue, citecolor=blue, urlcolor=blue}

\makeatletter
\usepackage{color}
\usepackage{babel}
\usepackage{breakurl}
\usepackage{orcidlink}

\renewcommand{\fnum@figure}{FIG.~\thefigure}
\@ifundefined{showcaptionsetup}{}{
 \PassOptionsToPackage{caption=false}{subfig}}

\usepackage[caption=false]{subfig} % use to save caption from breaking when using subfig

\makeatother

\begin{document}
\title{Suppression of Bremsstrahlung losses from relativistic plasma with
energy cutoff}
\author{Vadim R. Munirov\orcidlink{0000-0001-6711-1272}}
\email{vmunirov@princeton.edu}

\affiliation{Department of Astrophysical Sciences, Princeton University, Princeton,
New Jersey 08540, USA}
\author{Nathaniel J. Fisch\orcidlink{0000-0002-0301-7380}}
\affiliation{Department of Astrophysical Sciences, Princeton University, Princeton,
New Jersey 08540, USA}
\date{v1: April 3, 2023; v2: April 25, 2023; PRE accepted: 17 May, 2023}
\begin{abstract}
We study the effects of redistributing superthermal electrons on Bremsstrahlung
radiation from hot relativistic plasma. We consider thermal and nonthermal
distribution of electrons with an energy cutoff in the phase space
and explore the impact of the energy cutoff on Bremsstrahlung losses.
We discover that the redistribution of the superthermal electrons
into lower energies reduces radiative losses, which is in contrast
to nonrelativistic plasma. Finally, we discuss the possible relevance
of our results for open magnetic field line configurations and prospects
of the aneutronic fusion based on proton-Boron11 (p-B11) fuel. 
\end{abstract}
\maketitle

\section{Introduction}

Bremsstrahlung emission emerge whenever a charged particle moves in
the Coulomb field of other charged particles. Bremsstrahlung emission
is one of the primary mechanism of radiative energy loss from plasma
and is mainly characterized by its differential cross section. The
relativistic differential cross section of Bremsstrahlung emission
has been derived in a seminal work of Bethe and Heitler~\citep{BetheHeitler1934,Heitler1947},
where quantum matrix elements in the Born approximation were calculated.
Other notable papers and review works include~\citep{Sommerfeld1931,Bethe1954,Davies1954,Karzas1961,Haug1975a,Haug1975b,Koch1959,Akhiezer1965,Blumenthal1970,Bekefi1966,Jauch1980,Haug2004book}.

Bremsstrahlung is used for diagnostics purposes~\citep{Stevens1985,Voss1992,Peysson1996,Chen2006,Chen2008,Meadowcroft2012,Swanson2018,Kagan2019,Kumar2023,Brown1971}
and as a source of X-ray production~\citep{Jarrott2014,Cheng2015,Huntington2018,Underwood2020};
it is present in laser-plasma interactions~\citep{Singh2021,Vyskocil2018}
and determines dynamics of fast runaway electrons~\citep{Bakhtiari2005,Embreus2016}.
Bremsstrahlung emission is plentiful in astrophysics and has been
extensively studied in this context~\citep{Karzas1961,Kellogg1975,Svensson1982,Chluba2020,Pradler2021a,Pradler2021b,Pradler2021c}.
It is responsible for X-ray production in galaxy clusters~\citep{Sarazin1986,Sarazin2000,Pearce2000}
and solar flares~\citep{Brown1971,Haug1975c,Holman2003,Kontar2007},
plays a role in the physics of cosmic microwave background distortions~\citep{Sunyaev1970,Chluba2012},
and can be an important emission process for plasma around compact
objects~\citep{Narayan1994,Yarza2020}.

An inverse process of Bremsstrahlung absorption is one of the main
mechanisms of laser energy transfer in inertial confinement fusion
experiments~\citep{Keefe1982,Craxton2015,FirouziFarrashbandi2020}.
Many features of inverse Bremsstrahlung have been investigated~\citep{Rand1964,Shima1975,Brysk1975,Friedland1979,Mora1982,Balescu1982,GarbanLabaune1982,Cauble1985,Skupsky1987,Tsytovich1995,Tsytovich1996,Tsallis1997,Kostyukov2001,Wierling2001,Berger2004,Avetissian2013,FirouziFarrashbandi2015,FirouziFarrashbandi2020b}.
For example, in Refs.~\citep{Pashinin1978,Munirov2017b} recoil effect
in the electron-ion Bremsstrahlung absorption was studied, while influence
of strong laser fields on Bremsstrahlung absorption was explored in
Refs.~\citep{Bunkin1966,Denisov1968,Schlessinger1979,Pert1995,Shvets1997,Balakin2001,Brantov2003}.
Bremsstrahlung absorption is also critical for the opacity of astrophysical
plasmas~\citep{Krolik1984,Mihajlov2015}, such as high-temperature
stellar plasma~\citep{Itoh1985,Dimitrijevic2018} and the intracluster
plasma~\citep{Nozawa1998,Itoh2000,Itoh2002}. 

Both Bremsstrahlung emission and absorption crucially depend on the
distribution function of the charged particles. The distribution function
can differ substantially from a thermal distribution either naturally
or intentionally \textendash{} through phase space engineering. In
astrophysical settings, Bremsstrahlung emission from nonthermal power-law
distribution of electrons is present in supernova remnants \citep{Uchiyama2002,Vink2008,Fang2008},
clusters of galaxies~\citep{Sarazin2000}, and solar flares~\citep{Brown1971,Holman2003,Kontar2007}.
In laboratory settings, Langdon~\citep{Langdon1980} showed that
nonlinear effects in inverse Bremsstrahlung absorption lead to a distortion
of the electron distribution function towards a super-Gaussian, which
decreases the effectiveness of the energy transfer from laser to plasma~\citep{Matte1988}.
Intense radiation can even affect the electron distribution function
leading to magnetogenesis effects~\citep{Dubroca2004,Munirov2017a,Munirov2019,Ochs2020}.
The distribution function can also exhibit a significant degree of
anisotropy, which in turn affects Bremsstrahlung emission~\citep{Shohet1968,Dermer1986,Ferrante2001,Massone2004,Oparin2020}.

Fusion based on proton-Boron11 (p-B11) fuel has always been seen as
a very attractive method for generating clean energy due to its aneutronic
nature~\citep{Dawson1983,Davidson1979,Wurzel2022}. Because of the
temperature dependence of the p-B11 reaction cross section, fusion
with this fuel source requires plasma having a relativistic temperature
on the order of hundreds $\textrm{keV}$. Such high temperature plasmas
of relativistic temperatures emit significant amounts of radiation
with synchrotron and Bremsstrahlung emission being the major loss
mechanisms. These obstacles were deemed fatal for the feasibility
of fusion devices utilizing p-B11 fuel~\citep{Rider1995,Rider1997,Nevins1998}.
However, recent research has shown that the p-B11 reaction cross sections
could be larger than previously thought~\citep{Sikora2016} and that
the redistribution of fusion power from electrons to protons through
alpha channeling~\citep{Fisch1992a,Fisch1992b,Hay2015} make the
economical p-B11 fusion energy potentially viable~\citep{Putvinski2019,Ochs_pB11_2022}.
This inspired revival of interest in p-B11 fusion~\citep{Putvinski2019,Ochs_pB11_2022,Kolmes2022}
including some recent experimental endeavors~\citep{Kurilenkov2021,Lerner2023,Magee2023}.
Besides a magnetically confined p-B11, there are also growing efforts
with laser based p-B11 fusion~\citep{Belyaev2005,Kouhi2011,Picciotto2014,Hora2015,Eliezer2016,Labaune2016,Hora2017,Ruhl2022a,Ruhl2022b}.

In regard to synchrotron radiation, it was recently shown in Ref.~\citep{MlodikMunirov2023}
that synchrotron radiation from relativistic plasma can be meaningfully
reduced by redistribution of superthermal electrons into lower energies,
introducing an effective energy cutoff. Such an effective cutoff in
the energy distribution of electrons can emerge in open magnetic field
line configurations, such as mirror machines and inertial electrostatic
confinement devices. Relativistic Bremsstrahlung has a certain important
feature, that there is a long increasing tail in the probability of
Bremsstrahlung for large electron energies. This implies that the
redistribution of high energy electrons into lower energies should
lead to a decrease in Bremsstrahlung emission, similar to the effect
seen for synchrotron radiation in Ref.~\citep{MlodikMunirov2023}.

In this paper, we show that it is indeed possible to suppress the
production of Bremsstrahlung radiation from relativistic plasma by
redistributing superthermal electrons into lower energies. We evaluate
the power density of Bremsstrahlung radiation emitted from plasma
with an energy cutoff for different temperatures and energy cutoff
parameters, and determine the reduction in emission compared to the
case of thermal plasma. Finally, we discuss the possible relevance
of the present study for the p-B11 based fusion devices.

\section{Formulation of the problem and results}

We consider relativistic plasma of electron density $n_{e}$ with
the electrons described by the Maxwell\textendash Jüttner distribution
with an energy cutoff: 
\begin{equation}
f_{e}\left(\mathbf{p}\right)=\begin{cases}
N_{\textrm{const}}\frac{e^{-\frac{\gamma}{\theta_{T_{e}}}}}{4\pi m_{e}^{3}c^{3}\theta_{T_{e}}K_{2}\left(1/\theta_{T_{e}}\right)}, & \gamma\leq\gamma_{\textrm{max}},\\
0, & \gamma>\gamma_{\textrm{max}}.
\end{cases}\label{eq:f_e}
\end{equation}

\noindent Here, $\theta_{T_{e}}=T_{e}/(m_{e}c^{2})$ is the electron
temperature in the units of electron rest mass, $\gamma=\varepsilon/(m_{e}c^{2})=\sqrt{1+p^{2}/(m_{e}^{2}c^{2})}$
is the total electron energy in the units of electron rest mass or
the Lorentz factor, $\gamma_{\textrm{max}}$ is the energy cutoff
parameter, and $K_{2}$ is the modified Bessel function of the second
kind. The normalization constant $N_{\textrm{const}}$ is determined
through $\int f\left(\mathbf{p}\right)d\mathbf{p}=1$, so that for
pure Maxwell\textendash Jüttner distribution without a cutoff ($\gamma_{\textrm{max}}=\infty$)
the normalization constant is equal to unity. 

\noindent The total electron density $n_{e}$ is kept fixed, i.e.,
we do not throw away the electrons but rather redistribute them.

Our goal is to determine the total power density of Bremsstrahlung
radiation emitted from such a plasma. Self-absorption of Bremsstrahlung
radiation is usually negligible for magnetic confinement plasma and
thus to calculate the radiative losses we will solely concentrate
on spontaneous emission.

While for nonrelativistic plasma it is mainly electron-ion Coulomb
collisions that contribute to emission, for relativistic plasma electron-electron
Bremsstrahlung becomes comparable or even exceeds electron-ion contribution
and must be taken into account~\citep{Chen1982,Nozawa2009}.

The effective expression for the Bremsstrahlung power density emitted
from thermal relativistic plasma with the Maxwell\textendash Jüttner
distribution was derived in Ref.~\citep{Svensson1982}:

\begin{multline}
P_{\textrm{Br}}\approx7.56\times10^{-11}n_{e}^{2}\sqrt{\theta_{T_{e}}}\left[Z_{\textrm{eff}}(1+1.78\theta_{T_{e}}^{1.34})\right.\\
+\left.2.12\theta_{T_{e}}\left(1+1.1\theta_{T_{e}}+\theta_{T_{e}}^{2}-1.25\theta_{T_{e}}^{2.5}\right)\right]\textrm{eV}\,\textrm{cm}^{3}/\textrm{s}.\label{eq:P_br}
\end{multline}

\noindent Here, $Z_{\textrm{eff}}$ is the effective ion charge and
the formula is valid for relativistic, but not ultrarelativistic plasmas,
up to $\theta_{T_{e}}\leq1$. The first term in Eq.~(\ref{eq:P_br})
proportional to $Z_{\textrm{eff}}$ comes from electron-ion Bremsstrahlung;
it has a nonrelativistic leading order of $\sqrt{\theta_{T_{e}}}$,
while the $1.78\theta_{T_{e}}^{1.34}$ term inside the first round
brackets is a correction to it due to relativistic effects. The second
term comes from electron-electron Bremsstrahlung; it has a nonrelativistic
leading order of $\theta_{T_{e}}^{1.5}$ with the $1.1\theta_{T_{e}}+\theta_{T_{e}}^{2}-1.25\theta_{T_{e}}^{2.5}$
term inside the round brackets being a relativistic correction.

Expression (\ref{eq:P_br}) was used in Refs.~\citep{Putvinski2019,Ochs_pB11_2022}
(note that in Ref.~\citep{Ochs_pB11_2022} the $\theta_{T_{e}}^{2}$
term is missing in the second term due to electron-electron Bremsstrahlung)
to evaluate the energy budget of the p-B11 based fusion systems and
can be considered as a benchmark.

In the next two subsections we calculate the emitted radiation from
relativistic plasma described by the cutoff electron distribution
(\ref{eq:f_e}) due to electron-ion (Sec. \ref{subsec:Electron-ion-Bremsstrahlung})
and electron-electron (Sec. \ref{subsec:Electron-electron-Bremsstrahlung})
Bremsstrahlung and compare it with the thermal result given by Eq.~(\ref{eq:P_br}).
We will see that there is a reduction in Bremsstrahlung losses as
a result of introducing the energy cutoff and evaluate it.

\noindent 
\begin{figure}
\includegraphics[width=0.9\columnwidth]{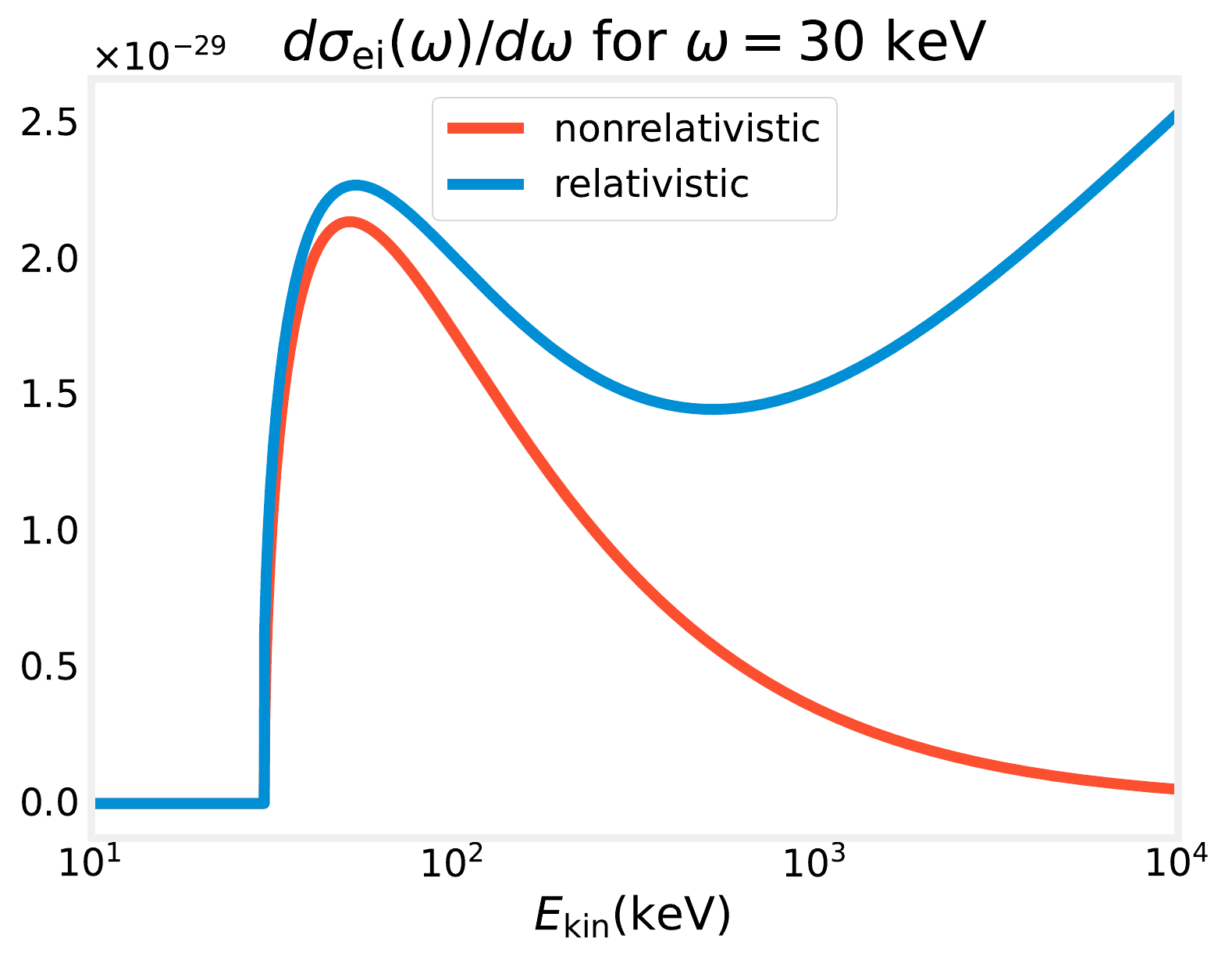}

\caption{\label{fig_dsigma_per_dw}The Elwert corrected relativistic Bethe-Heitler
differential cross section $d\sigma_{\textrm{ei}}\left(\omega\right)/d\omega$
{[}blue line given by Eq.~(\ref{eq:dsigma_BH}){]} in arbitrary units
for electron-ion Bremsstrahlung emission of $30\:\textrm{keV}$ photon
as a function of electron kinetic energy $\varepsilon_{\textrm{kin}}$.
The red line shows the nonrelativistic approximation of the Elwert
corrected Bethe-Heitler differential cross section (see Ref.~\citep{Jung1994}).}
\end{figure}

\noindent 
\begin{figure}[t]
\includegraphics[width=0.9\columnwidth]{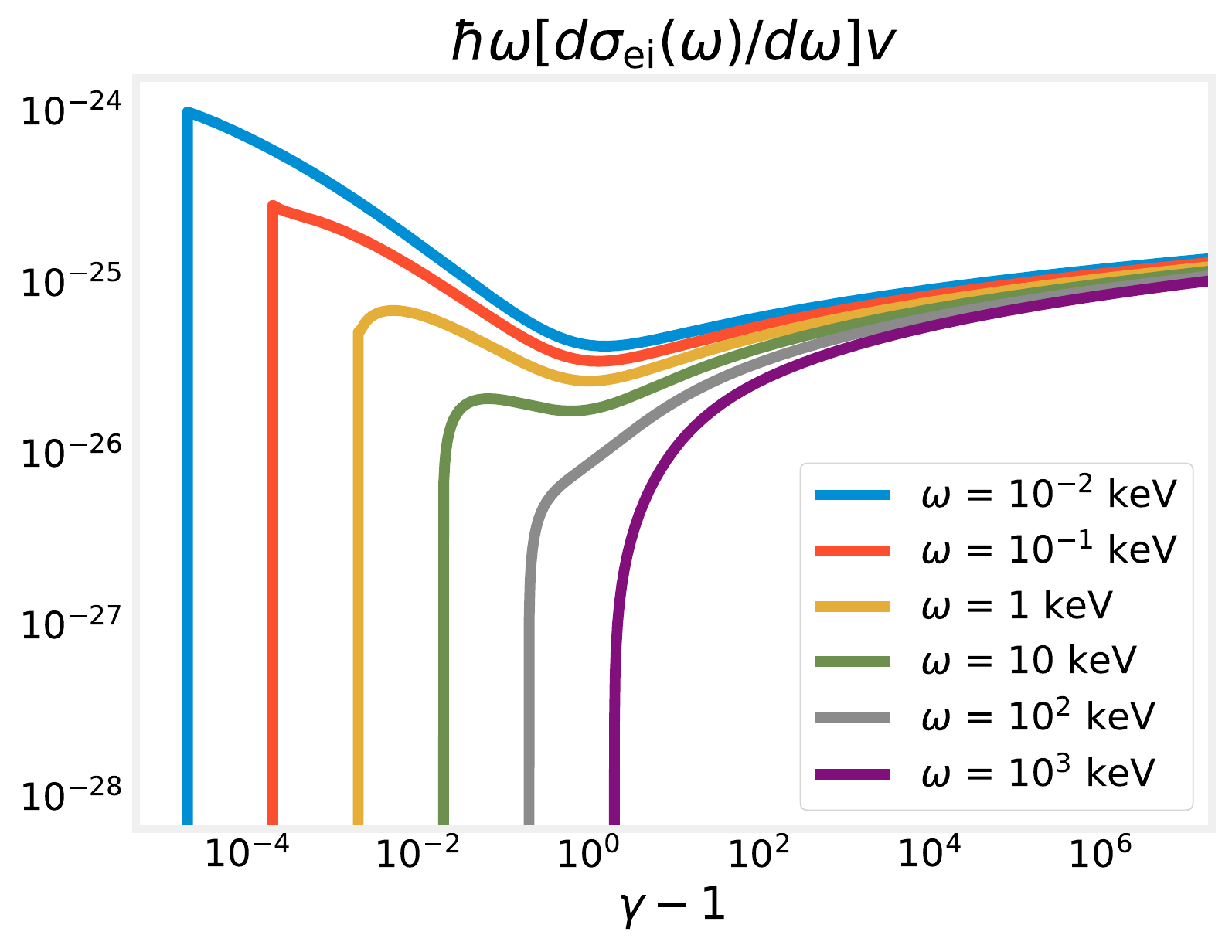}

\caption{\label{fig_w_dsigma_dw}The photon energy $\hbar\omega$ times the
Elwert corrected Bethe-Heitler differential cross section $d\sigma_{\textrm{ei}}\left(\omega\right)/d\omega$
{[}Eq.~(\ref{eq:dsigma_BH}){]} times the electron speed $v=pc^{2}/\varepsilon$
in arbitrary units as a function of the dimensionless electron kinetic
energy $\varepsilon_{\textrm{kin}}/(m_{e}c^{2})=\gamma-1$ for several
values of the photon energy. This product of the abovementioned three
quantities enters formula (\ref{eq:P_ei}) under the integral.}
\end{figure}

\subsection{Electron-ion Bremsstrahlung\label{subsec:Electron-ion-Bremsstrahlung}}

To calculate the radiative losses from relativistic plasma due to
Bremsstrahlung emission we need to know the corresponding differential
cross section. The relevant cross section for relativistic electron-ion
Bremsstrahlung is the Bethe-Heitler differential cross section~\citep{BetheHeitler1934,Heitler1947}.
It was used to derive expression~(\ref{eq:P_br}) and so Refs.~\citep{Putvinski2019,Ochs_pB11_2022}
also implicitly use it to calculate the thermal Bremsstrahlung losses
from the p-B11 plasma. The Bethe-Heitler differential cross section
for relativistic electron-ion Bremsstrahlung, including the Elwert
correction factor~\citep{Elwert1939}, is given by~\citep{Nozawa1998}

\begin{multline}
d\sigma_{\textrm{ei}}\left(\omega\right)=\alpha Z^{2}r_{e}^{2}\frac{p_{f}}{p}\frac{d\omega}{\omega}\frac{\eta_{f}}{\eta}\frac{1-e^{-2\pi\eta}}{1-e^{-2\pi\eta_{f}}}\\
\times\left\{ \frac{4}{3}-2\varepsilon\varepsilon_{f}\frac{p_{f}^{2}+p^{2}}{p_{f}^{2}p^{2}c^{2}}+m_{e}^{2}c^{2}\left(\frac{l_{f}\varepsilon}{p_{f}^{3}c}+\frac{l\varepsilon_{f}}{p^{3}c}-\frac{l_{f}l}{p_{f}p}\right)\right.\\
+L\left[\frac{8}{3}\frac{\varepsilon\varepsilon_{f}}{pp_{f}c^{2}}+\frac{\hbar^{2}\omega^{2}}{p^{3}p_{f}^{3}c^{6}}\left(\varepsilon^{2}\varepsilon_{f}^{2}+p^{2}p_{f}^{2}c^{4}\right)\right.\\
\left.\left.+\frac{m_{e}^{2}c^{2}\hbar\omega}{2pp_{f}}\left(\frac{\varepsilon\varepsilon_{f}+p^{2}c^{2}}{p^{3}c^{3}}l-\frac{\varepsilon\varepsilon_{f}+p_{f}^{2}c^{2}}{p_{f}^{3}c^{3}}l_{f}+\frac{2\hbar\omega\varepsilon\varepsilon_{f}}{p_{f}^{2}p^{2}c^{4}}\right)\right]\right\} .\label{eq:dsigma_BH}
\end{multline}

Equation (\ref{eq:dsigma_BH}) is valid when the Born approximation
is applicable, which requires $v/c=pc/\varepsilon\gg Z\alpha$. Here,
$\varepsilon$ is the electron energy, $\varepsilon_{f}$ is the electron
energy after emission of a photon, $p$ is the electron momentum,
$p_{f}$ is the electron momentum after emission of a photon, $\omega$
is the emitted photon angular frequency, $Z$ is the ion charge, $\alpha=e^{2}/(\hbar c)$
is the fine-structure constant, $r_{e}=e^{2}/(m_{e}c^{2})$ is the
classical electron radius, while

\begin{gather}
\varepsilon_{f}=\varepsilon-\hbar\omega,\\
l_{f}=2\ln\frac{\varepsilon_{f}+p_{f}c}{m_{e}c^{2}},l=2\ln\frac{\varepsilon+pc}{m_{e}c^{2}},\\
L=2\ln\frac{\varepsilon_{f}\varepsilon+p_{f}pc^{2}-m_{e}^{2}c^{4}}{m_{e}c^{2}\hbar\omega},\\
\eta_{f}=\frac{\alpha Z\varepsilon_{f}}{p_{f}c},\eta=\frac{\alpha Z\varepsilon}{pc}.
\end{gather}

\noindent 
\begin{figure}[t]
\includegraphics[width=0.9\columnwidth]{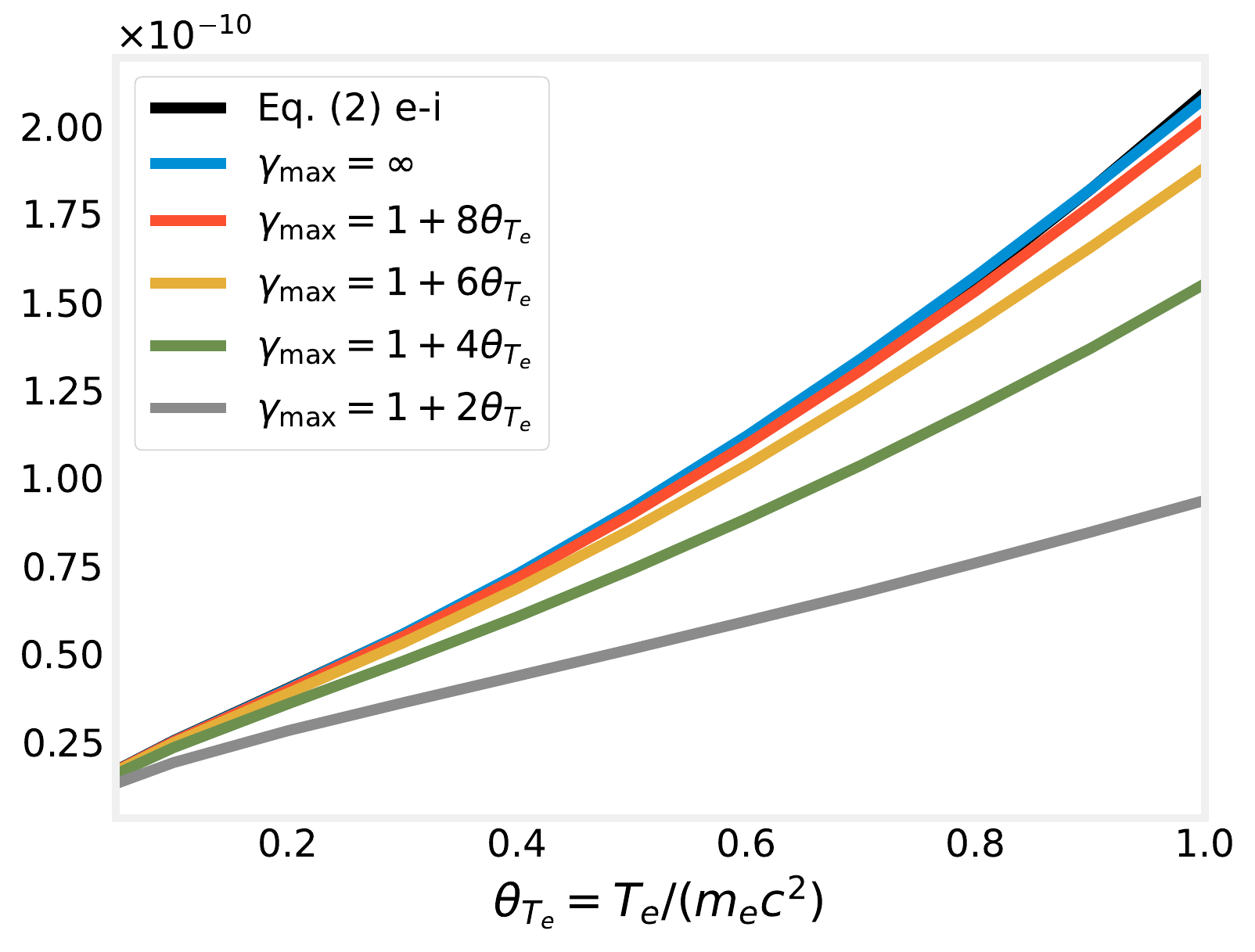}

\caption{\label{fig_abs_ei}Electron-ion Bremsstrahlung emitted power density
in arbitrary units as a function of the dimensionless electron temperature
$\theta_{T_{e}}$ for several values of the energy cutoff $\gamma_{\textrm{max}}$.}
\end{figure}

Figure~\ref{fig_dsigma_per_dw} shows the differential cross section
as a function of electron kinetic energy $\varepsilon_{\textrm{kin}}$.
We notice several important features from Fig.~\ref{fig_dsigma_per_dw}.
First, below $\varepsilon_{\textrm{kin}}=\hbar\omega$ the cross section
is zero, which is a manifestation of the fact that, due to energy
conservation, the electron cannot emit photon larger than its kinetic
energy. Second, we notice that after reaching a local maximum the
differential cross section does not decrease to zero for large values
of the kinetic energy but instead increases. This second feature is
of a relativistic nature and is not present in the nonrelativistic
calculations (see the red line in Fig.~\ref{fig_dsigma_per_dw} that
denotes the nonrelativistic approximation given in Ref.~\citep{Jung1994}).
It is this increased probability of Bremsstrahlung for large values
of the electron kinetic energy that is responsible for the additional
radiative losses in the relativistic regime.

The differential cross section~(\ref{eq:dsigma_BH}) together with
the electron distribution function determine the total electromagnetic
power density emitted from plasma due to electron-ion Bremsstrahlung:

\begin{equation}
P_{\textrm{ei}}=n_{e}n_{i}\iint\hbar\omega\frac{d\sigma_{\textrm{ei}}\left(\omega\right)}{d\omega}\frac{pc^{2}}{\varepsilon}f_{e}\left(\mathbf{p}\right)d\omega d\mathbf{p}.\label{eq:P_ei}
\end{equation}

Figure~\ref{fig_w_dsigma_dw} shows the product of the differential
cross section, the energy of the emitted photon, and the electron
speed that enters formula (\ref{eq:P_ei}) under the integral versus
the dimensionless electron kinetic energy $\varepsilon_{\textrm{kin}}/(m_{e}c^{2})=\gamma-1$
for a wide range of values of the emitted photon energy. We can see
that for large electron energies, the value of $\omega d\sigma_{\textrm{ei}}\left(\omega\right)/d\omega$
increases for all values of the photon energy.

The basic intuition that we extract from Figs.~\ref{fig_dsigma_per_dw}
and~\ref{fig_w_dsigma_dw} is that if we redistribute high energy
electrons into lower energies we could expect a reduction in the overall
emission. To check whether it is indeed correct we perform a series
of the numerical integrations for a range of electron temperatures
$T_{e}$ and the cutoff parameter $\gamma_{\textrm{max}}$.

\noindent 
\begin{figure}
\includegraphics[width=0.9\columnwidth]{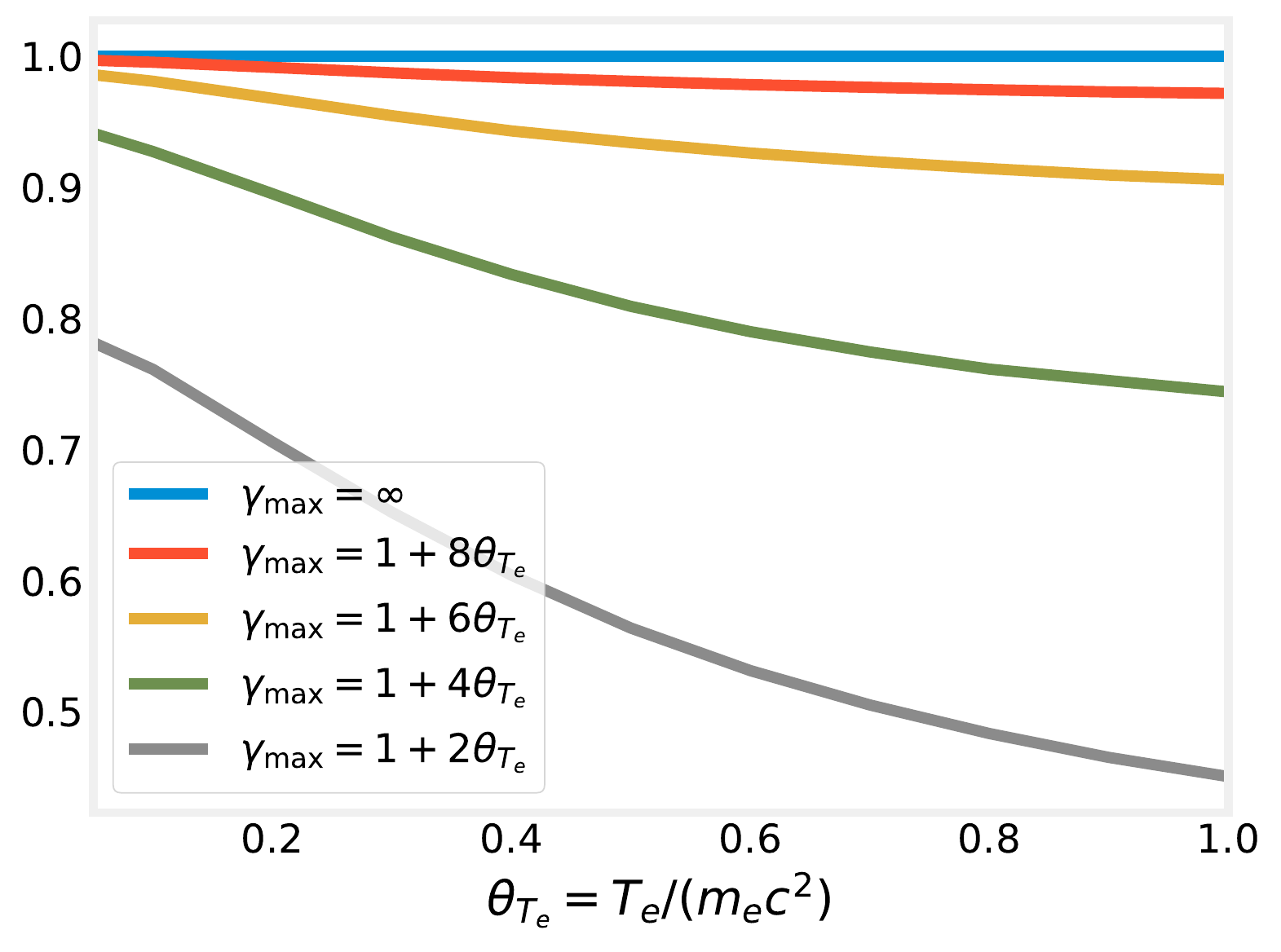}

\caption{\label{fig_rel_ei}Reduction in electron-ion Bremsstrahlung emission
relative to the thermal case as a function of the dimensionless electron
temperature $\theta_{T_{e}}$ for several values of the energy cutoff
$\gamma_{\textrm{max}}$. }
\end{figure}

The most important results of the paper are presented in Figs.~\ref{fig_abs_ei}
and~\ref{fig_rel_ei}. There we use Eqs.~(\ref{eq:dsigma_BH}) and~(\ref{eq:P_ei})
to calculate the power density due to electron-ion Bremsstrahlung
emission for different values of the electron temperature and the
energy cutoff. Figure~\ref{fig_abs_ei} shows the power density of
electron-ion Bremsstrahlung radiation generated by a plasma with the
electron distribution function that has an energy cutoff as well as
by a thermal plasma without a cutoff, both calculated numerically
and using a fitting formula of Eq.~(\ref{eq:P_br}); the graph shows
the corresponding curves for several values of the cutoff parameter
$\gamma_{\textrm{max}}$ versus the dimensionless electron temperature
$\theta_{T_{e}}=T_{e}/(m_{e}c^{2})$. As one would expect, we can
see that for thermal plasma without a cutoff ($\gamma_{\textrm{max}}=\infty$)
we reproduce the line given by Eq.~(\ref{eq:P_br}). Figure~\ref{fig_rel_ei}
is similar to Fig~\ref{fig_abs_ei}, but instead shows the power
density of Bremsstrahlung radiation from plasma with an energy cutoff
relative to the emission power from the thermal plasma. We can clearly
see the reduction in the emitted power for the distribution with an
energy cutoff. The larger the cutoff depth, the greater the reduction;
while for $\gamma_{\textrm{max}}\gtrsim1+8\theta_{T_{e}}$ the reduction
becomes negligible. We also see that as plasma becomes more relativistic,
i.e., $\theta_{T_{e}}$ approaches unity, the effect of the redistribution
becomes more pronounced.

Thus, we demonstrated that by redistributing electrons into lower
energies it is possible to mitigate electron-ion Bremsstrahlung emission
from relativistic plasma. Note that the opposite effect occurs, i.e.,
the Bremsstrahlung losses increase, for nonrelativistic plasma. This
is because in the nonrelativistic approximation, the differential
cross section decreases to zero for large electron energies, so it
is mainly thermal electrons that contribute to the emission. In the
relativistic case, the superthermal electrons contribute disproportionally
more to the emission and thus moving them into more thermal part of
the distribution reduces the overall radiative losses.

\subsection{Electron-electron Bremsstrahlung\label{subsec:Electron-electron-Bremsstrahlung}}

\noindent 
\begin{figure}
\includegraphics[width=0.9\columnwidth]{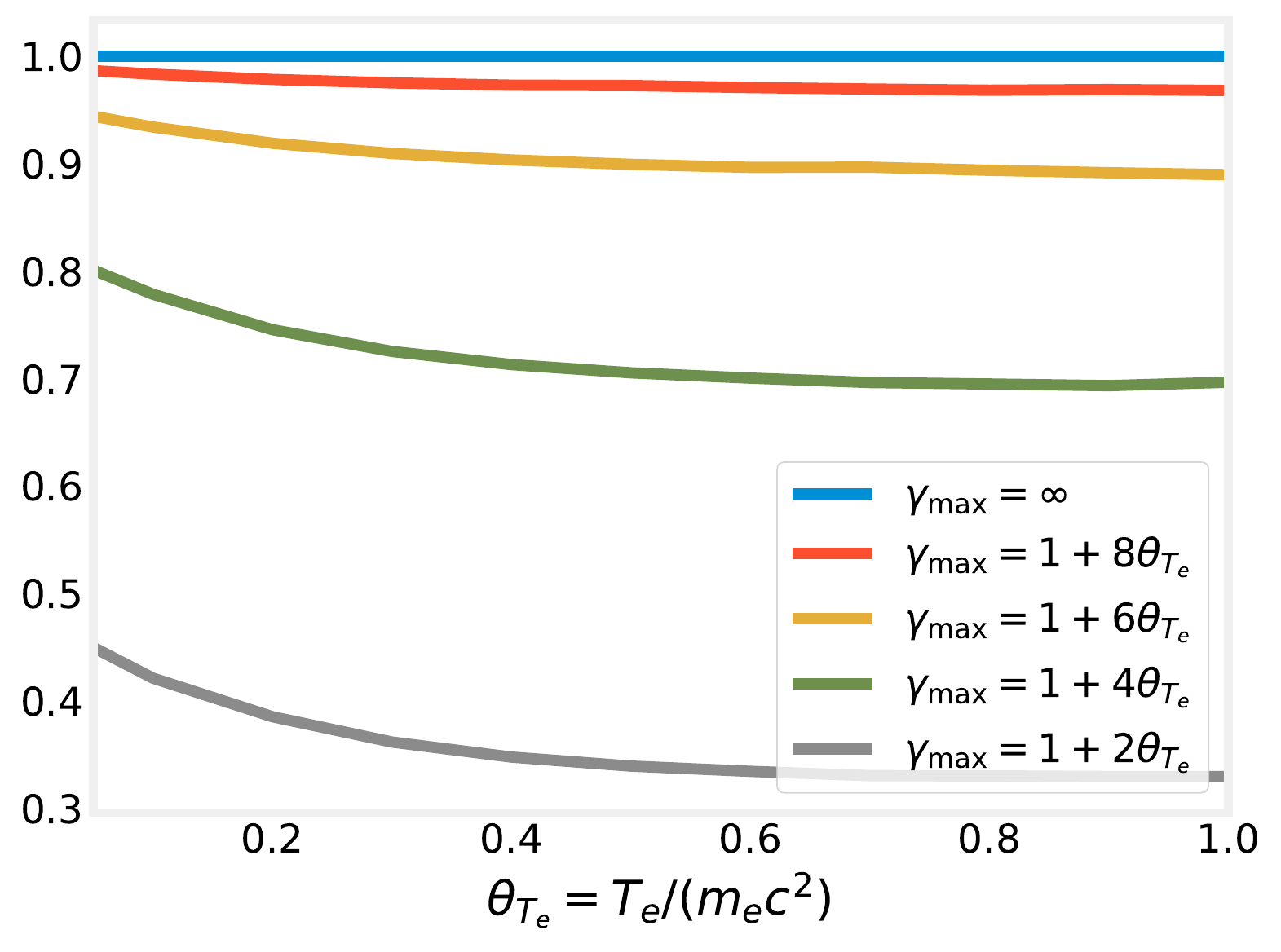}

\caption{\label{fig_rel_ee}Reduction in electron-electron Bremsstrahlung emission
relative to the thermal case as a function of the dimensionless electron
temperature $\theta_{T_{e}}$ for several values of the energy cutoff
$\gamma_{\textrm{max}}$. }
\end{figure}

We can expect that the redistribution of electrons into lower energies
will have an even greater impact on electron-electron Bremsstrahlung,
which has a quadrupole nature as opposed to a dipole nature of electron-ion
Bremsstrahlung~\citep{Maxon1967}. This is because the total electromagnetic
power density emitted from plasma due to electron-electron Bremsstrahlung
is obtained by integrating twice over the electron distribution~\citep{Haug1975b}:

\begin{multline}
P_{\textrm{ee}}=n_{e}^{2}m_{e}c^{3}\iint\frac{1}{2}\frac{\gamma_{1}+\gamma_{2}}{\gamma_{1}\gamma_{2}}\sqrt{\frac{1}{2}\left[\left(\mathbf{u}_{1}\cdotp\mathbf{u}_{2}\right)-1\right]}\\
\times\left(\int_{0}^{k_{\textrm{max}}}k_{\textrm{cm}}\frac{d\sigma}{dk_{\textrm{cm}}}dk_{\textrm{cm}}\right)f_{e}\left(\mathbf{u}_{1}\right)f_{e}\left(\mathbf{u}_{2}\right)d\mathbf{u}_{1}d\mathbf{u}_{2}.
\end{multline}

\noindent Here, $\mathbf{u}_{1}$, $\mathbf{u}_{2}$ are the dimensionless
momenta of two colliding electrons in the units of $m_{e}c$, i.e.,
$\mathbf{u}=\mathbf{p}/(m_{e}c)$, and $f_{e}\left(\mathbf{u}\right)$
is the electron distribution as a function of $\mathbf{u}$ properly
renormalized so that $\int f_{e}\left(\mathbf{u}\right)d\mathbf{u}=1$;
$k$ is the photon energy in the units of $m_{e}c^{2}$, $d\sigma/dk_{\textrm{cm}}$
is the differential cross section for electron-electron Bremsstrahlung,
the upper limit in the internal over $dk_{\textrm{cm}}$ is $k_{\textrm{max}}=u_{\textrm{cm}}^{2}/\gamma_{\textrm{cm}}$,
and the subscript ``$\textrm{cm}$'' refers to the center of mass
system of the colliding electrons; see Ref.~\citep{Haug1975b} for
the details.

The expression for the differential cross section of electron-electron
Bremsstrahlung is given in Refs.~\citep{Haug1975a,Haug1998}. However,
due to its complexity, a simplified treatment developed in Ref.~\citep{Haug1975b}
is commonly used. This simplified treatment was also employed in deriving
Eq.~(\ref{eq:P_br}) used in Refs.~ \citep{Svensson1982,Putvinski2019,OchsMunirov2023}.
For consistency and computational benefits we will also use this approach. 

In accordance with Ref.~\citep{Haug1975b}, we introduce the following
function:

\begin{multline}
J\left(\gamma_{1},\gamma_{2}\right)=\int_{\gamma_{1}\gamma_{2}-u_{1}u_{2}}^{\gamma_{1}\gamma_{2}+u_{1}u_{2}}\sqrt{\frac{1}{2}\left(\mu-1\right)}\\
\times\left(\int_{0}^{k_{\textrm{max}}}k_{\textrm{cm}}\frac{d\sigma}{dk_{\textrm{cm}}}dk_{\textrm{cm}}\right)d\mu,
\end{multline}

\noindent where $\mu=(\mathbf{u}_{1}\cdotp\mathbf{u}_{2})$ and $\int_{0}^{k_{\textrm{max}}}k_{\textrm{cm}}(d\sigma/dk_{\textrm{cm}})dk_{\textrm{cm}}$
is a function of $\mu$.

Then the emitted power density due to electron-electron Coulomb collisions
can be written as~\citep{Haug1975b}

\begin{equation}
P_{\textrm{ee}}=n_{e}^{2}m_{e}c^{3}\iint\frac{\gamma_{1}+\gamma_{2}}{4\gamma_{1}u_{1}\gamma_{2}u_{2}}J\left(\gamma_{1},\gamma_{2}\right)f_{e}\left(\gamma_{1}\right)f_{e}\left(\gamma_{2}\right)d\gamma_{1}d\gamma_{2},\label{eq:P_ee}
\end{equation}

\noindent where $f_{e}\left(\gamma\right)$ is the properly renormalized
electron distribution as a function of the electron energy in the
units of electron rest mass $\gamma$ so that $\int f_{e}\left(\gamma\right)d\gamma=1$.

It was demonstrated in Ref.~\citep{Haug1975b} that function $J\left(\gamma_{1},\gamma_{2}\right)$
can be approximated as

\begin{multline}
J\left(\gamma_{1},\gamma_{2}\right)=4\alpha r_{e}^{2}\Biggl\{\left(\frac{\mu}{2}-2\right)\sqrt{\mu^{2}-1}\\
-\frac{11}{12}\mu^{2}+\frac{20}{3}\mu-\frac{8}{3}\ln\left(\mu+1\right)\\
+\left[\frac{3}{2}+\left(\frac{\mu}{2}-\frac{8}{3}\frac{\mu+2}{\mu+1}\right)\sqrt{\mu^{2}-1}\right]\ln\left(\mu+\sqrt{\mu^{2}-1}\right)\\
\left.+\frac{7}{4}\ln^{2}\left(\mu+\sqrt{\mu^{2}-1}\right)\Biggr\}\right|_{\gamma_{1}\gamma_{2}-u_{1}u_{2}}^{\gamma_{1}\gamma_{2}+u_{1}u_{2}}.\label{eq:J_app}
\end{multline}

Using Eqs. ~(\ref{eq:P_ee}) and~(\ref{eq:J_app}) we can perform
a numerical integration to obtain the emitted power density for various
values of the electron temperature $T_{e}$ and the cutoff parameter
$\gamma_{\textrm{max}}$.

The reduction of the electron-electron Bremsstrahlung losses as a
function of the electron temperature for several values of the cutoff
parameter $\gamma_{\textrm{max}}$ relative to the power density of
the thermal Bremsstrahlung is shown in Fig.~\ref{fig_rel_ee}. We
can see that similar to the case of electron-ion Bremsstrahlung there
is a reduction in the radiative losses; in fact, the reduction even
exceeds that for the electron-ion case. 

\section{Possible relevance for aneutronic fusion}

\noindent 
\begin{figure}
\includegraphics[width=0.9\columnwidth]{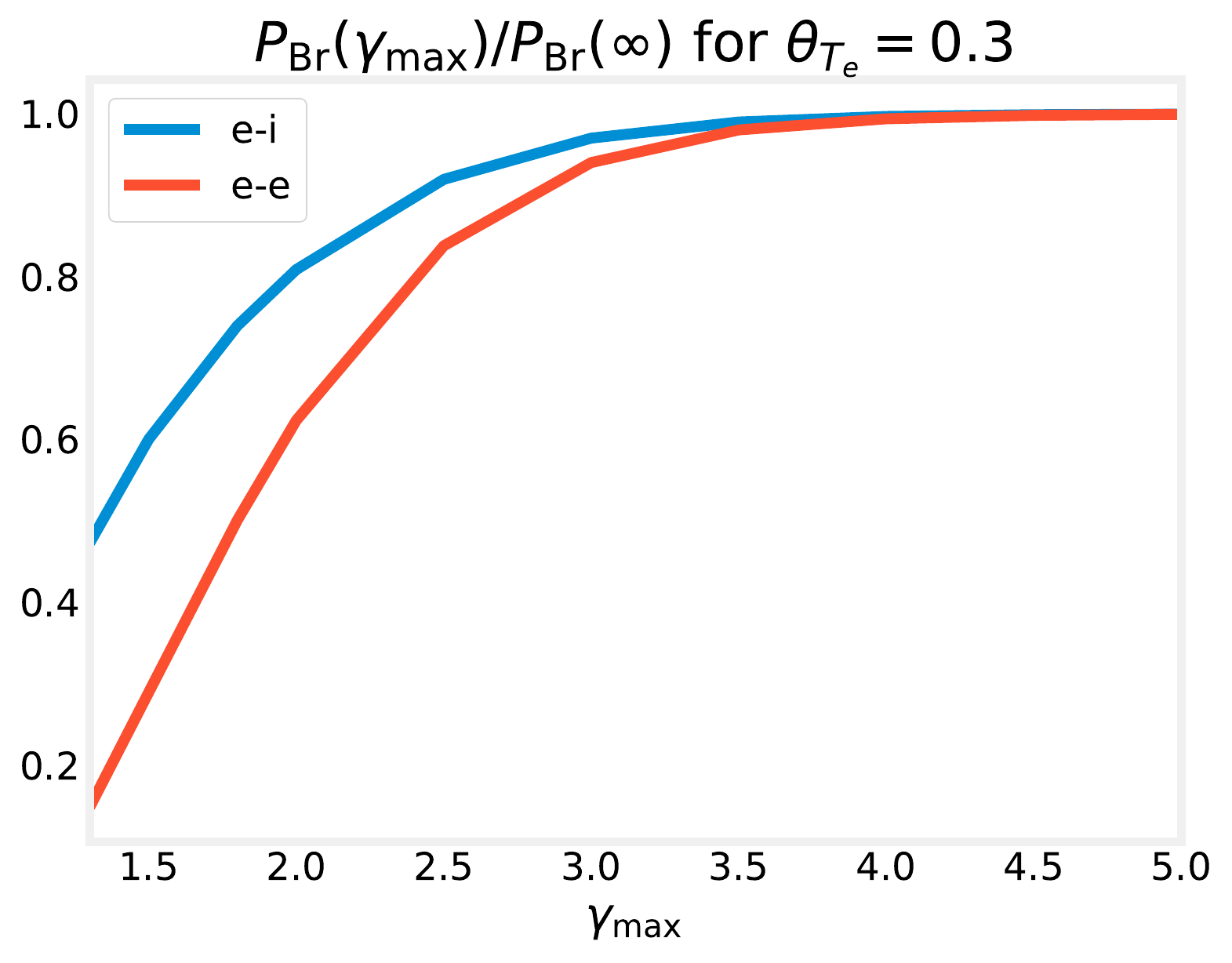}

\caption{\label{fig_pB11}Reduction in Bremsstrahlung emission relative to
the thermal case as a function the energy cutoff $\gamma_{\textrm{max}}$
for $\theta_{T_{e}}\approx0.3$ ($T_{e}\approx150\:\textrm{keV}$).}
\end{figure}

In this section, we speculate how the reduction in Bremsstrahlung
losses from relativistic plasma with an energy cutoff can potentially
help with the energy balance of p-B11 fusion reactors.

What has been envisioned as, in principle, a possible p-B11 fusion
reactor would operate at the typical electron temperature on the order
of $T_{e}\approx150\:\textrm{keV}$~\citep{Putvinski2019,Kolmes2022,Ochs_pB11_2022},
which corresponds to $\theta_{T_{e}}\approx0.3$. We plot the decrease
in Bremsstrahlung emission relative to the thermal case as a function
of the energy cutoff $\gamma_{\textrm{max}}$ for such a temperature
in Fig.~\ref{fig_pB11}. 

Open magnetic field line plasma devices, which are thought to be used
for the p-B11 based fusion, offer several methods for regulating the
confinement of electrons and ions. These methods include magnetic~\citep{Post1987,OchsMunirov2023},
electrostatic~\citep{Post1987,OchsMunirov2023}, centrifugal~\citep{Fetterman2011,White2018},
and ponderomotive~\citep{Rubin2023} confinement. For example, slow
electrons can be confined electrostatically, while fast electrons
are magnetically deconfined. This can be realized continuously: as
thermal electrons gain high energies through collisions and leave
the device, fast electrons are simultaneously removed, allowing electrostatic
forces, to which slow electrons are more sensitive, to replenish the
electron population to ensure charge neutrality.

Therefore, one possible way that the energy cutoff can emerge is due
to the electrostatic ambipolar potential. The typical electrostatic
ambipolar potential that can be established in such devices is roughly
estimated as $\left|e\varphi\right|\sim7T_{e}$~\citep{Post1987,OchsMunirov2023},
corresponding to the cutoff value $\gamma_{\textrm{max}}\approx2$.
We can see from Fig.~\ref{fig_pB11} that for $\gamma_{\textrm{max}}$
approximately $2$, there is a $20\%$ reduction in electron-ion and
a $40\%$ reduction in electron-electron Bremsstrahlung emission.
If we choose a more conservative value of $\gamma_{\textrm{max}}\approx2.5$,
we can observe a decrease of 10\% in electron-ion Bremsstrahlung emission
and a decrease of 20\% in electron-electron Bremsstrahlung emission.
Finally, as Fig.~\ref{fig_pB11} indicates, when the cutoff parameter
exceeds $\gamma_{\textrm{max}}\approx3.5$, the reduction in emission
becomes negligible. Thus, a reduction of at least 10\% can be potentially
achieved.

We caution, however, that removing the electron tail substitutes the
power loss in radiation with a power loss in kinetic energy. Nonetheless,
this effect can be advantageous in situations where it is much easier
to capture this energy in the form of fast electrons than potentially
damaging radiation coming all over the place. Bremsstrahlung energy
is very hard to capture in part because it is omnidirectional, requiring
any energy capture apparatus to completely surround the fusion device.
The power loss in electron kinetic energy is much more easily recovered
for several reasons. First of all, in a magnetic field, the electrons
are lost along the field lines, so it is highly localized and therefore
may be captured with localized devices. Second, it is in the form
of charged particle flow, for which energy capture can be efficient.
We also note that the distribution function in the mirror machines
may exhibit a significant anisotropy, so that the effective energy
cutoff for perpendicular and parallel energies can be noticeably different.
In addition, the emission of the radiation itself (both synchrotron
and Bremsstrahlung) will deplete the high energy tail and introduce
a natural energy cutoff. However, the total self-consistent picture
is outside of the scope of this work.

\section{Conclusion}

We considered Bremsstrahlung emission from a relativistic plasma of
fixed electron density but varying phase space distributions. We investigated
the impact of the energy cutoff on Bremsstrahlung losses and found
that for relativistic plasmas introducing energy cutoff through redistributing
superthermal electrons into lower energies can significantly decrease
radiative losses. This result is not entirely obvious even qualitatively,
let alone quantitatively; the nonrelativistic approximation of the
Bremsstrahlung effect could actually show an increase in the Bremsstrahlung
losses.

We conducted calculations to determine the potential reduction in
Bremsstrahlung emission for a typical p-B11 based fusion device. We
speculate that, if a number of assumptions are met, a meaningful reduction
of 10\% or even more can be potentially attained. Given the importance
of Bremsstrahlung losses for the operation of these systems, such
a potential reduction can help to relax the constraints on the energy
balance of the p-B11 based fusion. We caution, however, that such
strategy of trading radiation losses for kinetic losses must be accompanied
with a means of effective capture of the deconfined electrons, which
requires a certain design choices.

Finally, we highlight the potential relevance of our findings in astrophysical
scenarios. While it is power-law distributions that are typically
observed in astrophysical settings, under certain conditions, such
as formation of the effective magnetic mirror traps in the plasma
surrounding neutron stars and black holes, the distribution with an
energy cutoff can emerge, in which case the calculations presented
here will pertain. For example, it might change the Eddington limit
when it is influenced by Bremsstrahlung radiation~\citep{Beuermann1987,Ciotti2004}.

\section*{Acknowledgments}

This work was supported by ARPA-E Grant DE-AR0001554. 

\bibliographystyle{apsrev4-2}

\begin{thebibliography}{138}%
\makeatletter
\providecommand \@ifxundefined [1]{%
 \@ifx{#1\undefined}
}%
\providecommand \@ifnum [1]{%
 \ifnum #1\expandafter \@firstoftwo
 \else \expandafter \@secondoftwo
 \fi
}%
\providecommand \@ifx [1]{%
 \ifx #1\expandafter \@firstoftwo
 \else \expandafter \@secondoftwo
 \fi
}%
\providecommand \natexlab [1]{#1}%
\providecommand \enquote  [1]{``#1''}%
\providecommand \bibnamefont  [1]{#1}%
\providecommand \bibfnamefont [1]{#1}%
\providecommand \citenamefont [1]{#1}%
\providecommand \href@noop [0]{\@secondoftwo}%
\providecommand \href [0]{\begingroup \@sanitize@url \@href}%
\providecommand \@href[1]{\@@startlink{#1}\@@href}%
\providecommand \@@href[1]{\endgroup#1\@@endlink}%
\providecommand \@sanitize@url [0]{\catcode `\\12\catcode `\$12\catcode
  `\&12\catcode `\#12\catcode `\^12\catcode `\_12\catcode `\%12\relax}%
\providecommand \@@startlink[1]{}%
\providecommand \@@endlink[0]{}%
\providecommand \url  [0]{\begingroup\@sanitize@url \@url }%
\providecommand \@url [1]{\endgroup\@href {#1}{\urlprefix }}%
\providecommand \urlprefix  [0]{URL }%
\providecommand \Eprint [0]{\href }%
\providecommand \doibase [0]{https://doi.org/}%
\providecommand \selectlanguage [0]{\@gobble}%
\providecommand \bibinfo  [0]{\@secondoftwo}%
\providecommand \bibfield  [0]{\@secondoftwo}%
\providecommand \translation [1]{[#1]}%
\providecommand \BibitemOpen [0]{}%
\providecommand \bibitemStop [0]{}%
\providecommand \bibitemNoStop [0]{.\EOS\space}%
\providecommand \EOS [0]{\spacefactor3000\relax}%
\providecommand \BibitemShut  [1]{\csname bibitem#1\endcsname}%
\let\auto@bib@innerbib\@empty
%</preamble>
\bibitem [{\citenamefont {Bethe}\ and\ \citenamefont
  {Heitler}(1934)}]{BetheHeitler1934}%
  \BibitemOpen
  \bibfield  {author} {\bibinfo {author} {\bibfnamefont {H.}~\bibnamefont
  {Bethe}}\ and\ \bibinfo {author} {\bibfnamefont {W.}~\bibnamefont
  {Heitler}},\ }\href {https://doi.org/10.1098/rspa.1934.0140} {\bibfield
  {journal} {\bibinfo  {journal} {Proc. R. Soc. A}\ }\textbf {\bibinfo {volume}
  {146}},\ \bibinfo {pages} {83} (\bibinfo {year} {1934})}\BibitemShut
  {NoStop}%
\bibitem [{\citenamefont {Heitler}(1947)}]{Heitler1947}%
  \BibitemOpen
  \bibfield  {author} {\bibinfo {author} {\bibfnamefont {W.}~\bibnamefont
  {Heitler}},\ }\href@noop {} {\emph {\bibinfo {title} {The {Q}uantum {T}heory
  of {R}adiation}}},\ \bibinfo {edition} {2nd}\ ed.\ (\bibinfo  {publisher}
  {Oxford University Press},\ \bibinfo {address} {London},\ \bibinfo {year}
  {1947})\BibitemShut {NoStop}%
\bibitem [{\citenamefont {Sommerfeld}(1931)}]{Sommerfeld1931}%
  \BibitemOpen
  \bibfield  {author} {\bibinfo {author} {\bibfnamefont {A.}~\bibnamefont
  {Sommerfeld}},\ }\href
  {https://doi.org/https://doi.org/10.1002/andp.19314030302} {\bibfield
  {journal} {\bibinfo  {journal} {Ann. Phys. (Berlin)}\ }\textbf {\bibinfo
  {volume} {403}},\ \bibinfo {pages} {257} (\bibinfo {year}
  {1931})}\BibitemShut {NoStop}%
\bibitem [{\citenamefont {Bethe}\ and\ \citenamefont
  {Maximon}(1954)}]{Bethe1954}%
  \BibitemOpen
  \bibfield  {author} {\bibinfo {author} {\bibfnamefont {H.~A.}\ \bibnamefont
  {Bethe}}\ and\ \bibinfo {author} {\bibfnamefont {L.~C.}\ \bibnamefont
  {Maximon}},\ }\href {https://doi.org/10.1103/PhysRev.93.768} {\bibfield
  {journal} {\bibinfo  {journal} {Phys. Rev.}\ }\textbf {\bibinfo {volume}
  {93}},\ \bibinfo {pages} {768} (\bibinfo {year} {1954})}\BibitemShut
  {NoStop}%
\bibitem [{\citenamefont {Davies}\ \emph {et~al.}(1954)\citenamefont {Davies},
  \citenamefont {Bethe},\ and\ \citenamefont {Maximon}}]{Davies1954}%
  \BibitemOpen
  \bibfield  {author} {\bibinfo {author} {\bibfnamefont {H.}~\bibnamefont
  {Davies}}, \bibinfo {author} {\bibfnamefont {H.~A.}\ \bibnamefont {Bethe}},\
  and\ \bibinfo {author} {\bibfnamefont {L.~C.}\ \bibnamefont {Maximon}},\
  }\href {https://doi.org/10.1103/PhysRev.93.788} {\bibfield  {journal}
  {\bibinfo  {journal} {Phys. Rev.}\ }\textbf {\bibinfo {volume} {93}},\
  \bibinfo {pages} {788} (\bibinfo {year} {1954})}\BibitemShut {NoStop}%
\bibitem [{\citenamefont {{Karzas}}\ and\ \citenamefont
  {{Latter}}(1961)}]{Karzas1961}%
  \BibitemOpen
  \bibfield  {author} {\bibinfo {author} {\bibfnamefont {W.~J.}\ \bibnamefont
  {{Karzas}}}\ and\ \bibinfo {author} {\bibfnamefont {R.}~\bibnamefont
  {{Latter}}},\ }\href {https://doi.org/10.1086/190063} {\bibfield  {journal}
  {\bibinfo  {journal} {Astrophys. J., Suppl. Ser.}\ }\textbf {\bibinfo
  {volume} {6}},\ \bibinfo {pages} {167} (\bibinfo {year} {1961})}\BibitemShut
  {NoStop}%
\bibitem [{\citenamefont {Haug}(1975{\natexlab{a}})}]{Haug1975a}%
  \BibitemOpen
  \bibfield  {author} {\bibinfo {author} {\bibfnamefont {E.}~\bibnamefont
  {Haug}},\ }\href {https://doi.org/doi:10.1515/zna-1975-0901} {\bibfield
  {journal} {\bibinfo  {journal} {Z. Naturforsch. A}\ }\textbf {\bibinfo
  {volume} {30}},\ \bibinfo {pages} {1099} (\bibinfo {year}
  {1975}{\natexlab{a}})}\BibitemShut {NoStop}%
\bibitem [{\citenamefont {Haug}(1975{\natexlab{b}})}]{Haug1975b}%
  \BibitemOpen
  \bibfield  {author} {\bibinfo {author} {\bibfnamefont {E.}~\bibnamefont
  {Haug}},\ }\href {https://doi.org/doi:10.1515/zna-1975-1206} {\bibfield
  {journal} {\bibinfo  {journal} {Z. Naturforsch. A}\ }\textbf {\bibinfo
  {volume} {30}},\ \bibinfo {pages} {1546} (\bibinfo {year}
  {1975}{\natexlab{b}})}\BibitemShut {NoStop}%
\bibitem [{\citenamefont {Koch}\ and\ \citenamefont {Motz}(1959)}]{Koch1959}%
  \BibitemOpen
  \bibfield  {author} {\bibinfo {author} {\bibfnamefont {H.~W.}\ \bibnamefont
  {Koch}}\ and\ \bibinfo {author} {\bibfnamefont {J.~W.}\ \bibnamefont
  {Motz}},\ }\href {https://doi.org/10.1103/RevModPhys.31.920} {\bibfield
  {journal} {\bibinfo  {journal} {Rev. Mod. Phys.}\ }\textbf {\bibinfo {volume}
  {31}},\ \bibinfo {pages} {920} (\bibinfo {year} {1959})}\BibitemShut
  {NoStop}%
\bibitem [{\citenamefont {Akhiezer}\ and\ \citenamefont
  {Berestetskii}(1965)}]{Akhiezer1965}%
  \BibitemOpen
  \bibfield  {author} {\bibinfo {author} {\bibfnamefont {A.~I.}\ \bibnamefont
  {Akhiezer}}\ and\ \bibinfo {author} {\bibfnamefont {V.~B.}\ \bibnamefont
  {Berestetskii}},\ }\href@noop {} {\emph {\bibinfo {title} {Quantum
  Electrodynamics}}},\ Monographs on Physics \& Astronomical\ (\bibinfo
  {publisher} {John Wiley \& Sons},\ \bibinfo {address} {Nashville, TN},\
  \bibinfo {year} {1965})\BibitemShut {NoStop}%
\bibitem [{\citenamefont {Blumenthal}\ and\ \citenamefont
  {Gould}(1970)}]{Blumenthal1970}%
  \BibitemOpen
  \bibfield  {author} {\bibinfo {author} {\bibfnamefont {G.~R.}\ \bibnamefont
  {Blumenthal}}\ and\ \bibinfo {author} {\bibfnamefont {R.~J.}\ \bibnamefont
  {Gould}},\ }\href {https://doi.org/10.1103/RevModPhys.42.237} {\bibfield
  {journal} {\bibinfo  {journal} {Rev. Mod. Phys.}\ }\textbf {\bibinfo {volume}
  {42}},\ \bibinfo {pages} {237} (\bibinfo {year} {1970})}\BibitemShut
  {NoStop}%
\bibitem [{\citenamefont {Bekefi}(1966)}]{Bekefi1966}%
  \BibitemOpen
  \bibfield  {author} {\bibinfo {author} {\bibfnamefont {G.}~\bibnamefont
  {Bekefi}},\ }\href@noop {} {\emph {\bibinfo {title} {{R}adiation {P}rocesses
  in {P}lasmas}}}\ (\bibinfo  {publisher} {Wiley},\ \bibinfo {address} {New
  York},\ \bibinfo {year} {1966})\BibitemShut {NoStop}%
\bibitem [{\citenamefont {Jauch}\ and\ \citenamefont
  {Rohrlich}(1980)}]{Jauch1980}%
  \BibitemOpen
  \bibfield  {author} {\bibinfo {author} {\bibfnamefont {J.~M.}\ \bibnamefont
  {Jauch}}\ and\ \bibinfo {author} {\bibfnamefont {F.}~\bibnamefont
  {Rohrlich}},\ }\href@noop {} {\emph {\bibinfo {title} {The {T}heory of
  {P}hotons and {E}lectrons}}},\ \bibinfo {edition} {2nd}\ ed.,\ Theoretical
  and Mathematical Physics\ (\bibinfo  {publisher} {Springer},\ \bibinfo
  {address} {Berlin, Germany},\ \bibinfo {year} {1980})\BibitemShut {NoStop}%
\bibitem [{\citenamefont {Haug}\ and\ \citenamefont
  {Nakel}(2004)}]{Haug2004book}%
  \BibitemOpen
  \bibfield  {author} {\bibinfo {author} {\bibfnamefont {E.}~\bibnamefont
  {Haug}}\ and\ \bibinfo {author} {\bibfnamefont {W.}~\bibnamefont {Nakel}},\
  }\href {https://doi.org/10.1142/5371} {\emph {\bibinfo {title} {The
  {E}lementary {P}rocess of {B}remsstrahlung}}}\ (\bibinfo  {publisher} {World
  Scientific},\ \bibinfo {address} {Singapore},\ \bibinfo {year}
  {2004})\BibitemShut {NoStop}%
\bibitem [{\citenamefont {Stevens}\ \emph {et~al.}(1985)\citenamefont
  {Stevens}, \citenamefont {Goeler}, \citenamefont {Bernabei}, \citenamefont
  {Bitter}, \citenamefont {Chu}, \citenamefont {Efthimion}, \citenamefont
  {Fisch}, \citenamefont {Hooke}, \citenamefont {Hosea}, \citenamefont {Jobes},
  \citenamefont {Karney}, \citenamefont {Meservey}, \citenamefont {Motley},\
  and\ \citenamefont {Taylor}}]{Stevens1985}%
  \BibitemOpen
  \bibfield  {author} {\bibinfo {author} {\bibfnamefont {J.}~\bibnamefont
  {Stevens}}, \bibinfo {author} {\bibfnamefont {S.~V.}\ \bibnamefont {Goeler}},
  \bibinfo {author} {\bibfnamefont {S.}~\bibnamefont {Bernabei}}, \bibinfo
  {author} {\bibfnamefont {M.}~\bibnamefont {Bitter}}, \bibinfo {author}
  {\bibfnamefont {T.}~\bibnamefont {Chu}}, \bibinfo {author} {\bibfnamefont
  {P.}~\bibnamefont {Efthimion}}, \bibinfo {author} {\bibfnamefont
  {N.}~\bibnamefont {Fisch}}, \bibinfo {author} {\bibfnamefont
  {W.}~\bibnamefont {Hooke}}, \bibinfo {author} {\bibfnamefont
  {J.}~\bibnamefont {Hosea}}, \bibinfo {author} {\bibfnamefont
  {F.}~\bibnamefont {Jobes}}, \bibinfo {author} {\bibfnamefont
  {C.}~\bibnamefont {Karney}}, \bibinfo {author} {\bibfnamefont
  {E.}~\bibnamefont {Meservey}}, \bibinfo {author} {\bibfnamefont
  {R.}~\bibnamefont {Motley}},\ and\ \bibinfo {author} {\bibfnamefont
  {G.}~\bibnamefont {Taylor}},\ }\href
  {https://doi.org/10.1088/0029-5515/25/11/002} {\bibfield  {journal} {\bibinfo
   {journal} {Nucl. Fusion}\ }\textbf {\bibinfo {volume} {25}},\ \bibinfo
  {pages} {1529} (\bibinfo {year} {1985})}\BibitemShut {NoStop}%
\bibitem [{\citenamefont {Voss}\ and\ \citenamefont {Fisch}(1992)}]{Voss1992}%
  \BibitemOpen
  \bibfield  {author} {\bibinfo {author} {\bibfnamefont {K.~E.}\ \bibnamefont
  {Voss}}\ and\ \bibinfo {author} {\bibfnamefont {N.~J.}\ \bibnamefont
  {Fisch}},\ }\href {https://doi.org/10.1063/1.860221} {\bibfield  {journal}
  {\bibinfo  {journal} {Phys. Fluids B}\ }\textbf {\bibinfo {volume} {4}},\
  \bibinfo {pages} {762} (\bibinfo {year} {1992})}\BibitemShut {NoStop}%
\bibitem [{\citenamefont {Peysson}\ and\ \citenamefont
  {Arslanbekov}(1996)}]{Peysson1996}%
  \BibitemOpen
  \bibfield  {author} {\bibinfo {author} {\bibfnamefont {Y.}~\bibnamefont
  {Peysson}}\ and\ \bibinfo {author} {\bibfnamefont {R.}~\bibnamefont
  {Arslanbekov}},\ }\href
  {https://doi.org/https://doi.org/10.1016/S0168-9002(96)00316-6} {\bibfield
  {journal} {\bibinfo  {journal} {Nucl. Instrum. Methods Phys. Res.}\ }\textbf
  {\bibinfo {volume} {380}},\ \bibinfo {pages} {423} (\bibinfo {year}
  {1996})}\BibitemShut {NoStop}%
\bibitem [{\citenamefont {Chen}\ \emph {et~al.}(2006)\citenamefont {Chen},
  \citenamefont {Wan}, \citenamefont {Lin}, \citenamefont {Shi}, \citenamefont
  {Hu},\ and\ \citenamefont {Ma}}]{Chen2006}%
  \BibitemOpen
  \bibfield  {author} {\bibinfo {author} {\bibfnamefont {Z.}~\bibnamefont
  {Chen}}, \bibinfo {author} {\bibfnamefont {B.}~\bibnamefont {Wan}}, \bibinfo
  {author} {\bibfnamefont {S.}~\bibnamefont {Lin}}, \bibinfo {author}
  {\bibfnamefont {Y.}~\bibnamefont {Shi}}, \bibinfo {author} {\bibfnamefont
  {L.}~\bibnamefont {Hu}},\ and\ \bibinfo {author} {\bibfnamefont
  {T.}~\bibnamefont {Ma}},\ }\href
  {https://doi.org/https://doi.org/10.1016/j.nima.2005.12.200} {\bibfield
  {journal} {\bibinfo  {journal} {Nucl. Instrum. Methods Phys. Res.}\ }\textbf
  {\bibinfo {volume} {560}},\ \bibinfo {pages} {558} (\bibinfo {year}
  {2006})}\BibitemShut {NoStop}%
\bibitem [{\citenamefont {Chen}\ \emph {et~al.}(2008)\citenamefont {Chen},
  \citenamefont {King}, \citenamefont {Key}, \citenamefont {Akli},
  \citenamefont {Beg}, \citenamefont {Chen}, \citenamefont {Freeman},
  \citenamefont {Link}, \citenamefont {Mackinnon}, \citenamefont {MacPhee},
  \citenamefont {Patel}, \citenamefont {Porkolab}, \citenamefont {Stephens},\
  and\ \citenamefont {Van~Woerkom}}]{Chen2008}%
  \BibitemOpen
  \bibfield  {author} {\bibinfo {author} {\bibfnamefont {C.~D.}\ \bibnamefont
  {Chen}}, \bibinfo {author} {\bibfnamefont {J.~A.}\ \bibnamefont {King}},
  \bibinfo {author} {\bibfnamefont {M.~H.}\ \bibnamefont {Key}}, \bibinfo
  {author} {\bibfnamefont {K.~U.}\ \bibnamefont {Akli}}, \bibinfo {author}
  {\bibfnamefont {F.~N.}\ \bibnamefont {Beg}}, \bibinfo {author} {\bibfnamefont
  {H.}~\bibnamefont {Chen}}, \bibinfo {author} {\bibfnamefont {R.~R.}\
  \bibnamefont {Freeman}}, \bibinfo {author} {\bibfnamefont {A.}~\bibnamefont
  {Link}}, \bibinfo {author} {\bibfnamefont {A.~J.}\ \bibnamefont {Mackinnon}},
  \bibinfo {author} {\bibfnamefont {A.~G.}\ \bibnamefont {MacPhee}}, \bibinfo
  {author} {\bibfnamefont {P.~K.}\ \bibnamefont {Patel}}, \bibinfo {author}
  {\bibfnamefont {M.}~\bibnamefont {Porkolab}}, \bibinfo {author}
  {\bibfnamefont {R.~B.}\ \bibnamefont {Stephens}},\ and\ \bibinfo {author}
  {\bibfnamefont {L.~D.}\ \bibnamefont {Van~Woerkom}},\ }\href
  {https://doi.org/10.1063/1.2964231} {\bibfield  {journal} {\bibinfo
  {journal} {Rev. Sci. Instrum.}\ }\textbf {\bibinfo {volume} {79}},\ \bibinfo
  {pages} {10E305} (\bibinfo {year} {2008})}\BibitemShut {NoStop}%
\bibitem [{\citenamefont {Meadowcroft}\ and\ \citenamefont
  {Edwards}(2012)}]{Meadowcroft2012}%
  \BibitemOpen
  \bibfield  {author} {\bibinfo {author} {\bibfnamefont {A.~L.}\ \bibnamefont
  {Meadowcroft}}\ and\ \bibinfo {author} {\bibfnamefont {R.~D.}\ \bibnamefont
  {Edwards}},\ }\href {https://doi.org/10.1109/TPS.2012.2201175} {\bibfield
  {journal} {\bibinfo  {journal} {IEEE Trans. Plasma Sci.}\ }\textbf {\bibinfo
  {volume} {40}},\ \bibinfo {pages} {1992} (\bibinfo {year}
  {2012})}\BibitemShut {NoStop}%
\bibitem [{\citenamefont {Swanson}\ \emph {et~al.}(2018)\citenamefont
  {Swanson}, \citenamefont {Jandovitz},\ and\ \citenamefont
  {Cohen}}]{Swanson2018}%
  \BibitemOpen
  \bibfield  {author} {\bibinfo {author} {\bibfnamefont {C.}~\bibnamefont
  {Swanson}}, \bibinfo {author} {\bibfnamefont {P.}~\bibnamefont {Jandovitz}},\
  and\ \bibinfo {author} {\bibfnamefont {S.~A.}\ \bibnamefont {Cohen}},\ }\href
  {https://doi.org/10.1063/1.5019572} {\bibfield  {journal} {\bibinfo
  {journal} {AIP Adv.}\ }\textbf {\bibinfo {volume} {8}},\ \bibinfo {pages}
  {025222} (\bibinfo {year} {2018})}\BibitemShut {NoStop}%
\bibitem [{\citenamefont {Kagan}\ \emph {et~al.}(2019)\citenamefont {Kagan},
  \citenamefont {Landen}, \citenamefont {Svyatskiy}, \citenamefont {Sio},
  \citenamefont {Kabadi}, \citenamefont {Simpson}, \citenamefont
  {Gatu~Johnson}, \citenamefont {Frenje}, \citenamefont {Petrasso},
  \citenamefont {Shah}, \citenamefont {Joshi}, \citenamefont {Hakel},
  \citenamefont {Weber}, \citenamefont {Rinderknecht}, \citenamefont {Thorn},
  \citenamefont {Schneider}, \citenamefont {Bradley},\ and\ \citenamefont
  {Kilkenny}}]{Kagan2019}%
  \BibitemOpen
  \bibfield  {author} {\bibinfo {author} {\bibfnamefont {G.}~\bibnamefont
  {Kagan}}, \bibinfo {author} {\bibfnamefont {O.}~\bibnamefont {Landen}},
  \bibinfo {author} {\bibfnamefont {D.}~\bibnamefont {Svyatskiy}}, \bibinfo
  {author} {\bibfnamefont {H.}~\bibnamefont {Sio}}, \bibinfo {author}
  {\bibfnamefont {N.}~\bibnamefont {Kabadi}}, \bibinfo {author} {\bibfnamefont
  {R.}~\bibnamefont {Simpson}}, \bibinfo {author} {\bibfnamefont
  {M.}~\bibnamefont {Gatu~Johnson}}, \bibinfo {author} {\bibfnamefont
  {J.}~\bibnamefont {Frenje}}, \bibinfo {author} {\bibfnamefont
  {R.}~\bibnamefont {Petrasso}}, \bibinfo {author} {\bibfnamefont
  {R.}~\bibnamefont {Shah}}, \bibinfo {author} {\bibfnamefont {T.}~\bibnamefont
  {Joshi}}, \bibinfo {author} {\bibfnamefont {P.}~\bibnamefont {Hakel}},
  \bibinfo {author} {\bibfnamefont {T.}~\bibnamefont {Weber}}, \bibinfo
  {author} {\bibfnamefont {H.}~\bibnamefont {Rinderknecht}}, \bibinfo {author}
  {\bibfnamefont {D.}~\bibnamefont {Thorn}}, \bibinfo {author} {\bibfnamefont
  {M.}~\bibnamefont {Schneider}}, \bibinfo {author} {\bibfnamefont
  {D.}~\bibnamefont {Bradley}},\ and\ \bibinfo {author} {\bibfnamefont
  {J.}~\bibnamefont {Kilkenny}},\ }\href
  {https://doi.org/https://doi.org/10.1002/ctpp.201800078} {\bibfield
  {journal} {\bibinfo  {journal} {Contrib. to Plasma Phys.}\ }\textbf {\bibinfo
  {volume} {59}},\ \bibinfo {pages} {181} (\bibinfo {year} {2019})}\BibitemShut
  {NoStop}%
\bibitem [{\citenamefont {Kumar}\ \emph {et~al.}(2023)\citenamefont {Kumar},
  \citenamefont {Pandya},\ and\ \citenamefont {Sharma}}]{Kumar2023}%
  \BibitemOpen
  \bibfield  {author} {\bibinfo {author} {\bibfnamefont {J.}~\bibnamefont
  {Kumar}}, \bibinfo {author} {\bibfnamefont {S.~P.}\ \bibnamefont {Pandya}},\
  and\ \bibinfo {author} {\bibfnamefont {P.}~\bibnamefont {Sharma}},\ }\href
  {https://doi.org/10.1088/1748-0221/18/03/P03040} {\bibfield  {journal}
  {\bibinfo  {journal} {J. Instrum.}\ }\textbf {\bibinfo {volume} {18}}\bibinfo
   {number} { (03)},\ \bibinfo {pages} {P03040}}\BibitemShut {NoStop}%
\bibitem [{\citenamefont {Brown}(1971)}]{Brown1971}%
  \BibitemOpen
\bibfield  {number} {  }\bibfield  {author} {\bibinfo {author} {\bibfnamefont
  {J.~C.}\ \bibnamefont {Brown}},\ }\href {https://doi.org/10.1007/BF00149070}
  {\bibfield  {journal} {\bibinfo  {journal} {Sol. Phys.}\ }\textbf {\bibinfo
  {volume} {18}},\ \bibinfo {pages} {489} (\bibinfo {year} {1971})}\BibitemShut
  {NoStop}%
\bibitem [{\citenamefont {Jarrott}\ \emph {et~al.}(2014)\citenamefont
  {Jarrott}, \citenamefont {Kemp}, \citenamefont {Divol}, \citenamefont
  {Mariscal}, \citenamefont {Westover}, \citenamefont {McGuffey}, \citenamefont
  {Beg}, \citenamefont {Suggit}, \citenamefont {Chen}, \citenamefont {Hey},
  \citenamefont {Maddox}, \citenamefont {Hawreliak}, \citenamefont {Park},
  \citenamefont {Remington}, \citenamefont {Wei},\ and\ \citenamefont
  {MacPhee}}]{Jarrott2014}%
  \BibitemOpen
  \bibfield  {author} {\bibinfo {author} {\bibfnamefont {L.~C.}\ \bibnamefont
  {Jarrott}}, \bibinfo {author} {\bibfnamefont {A.~J.}\ \bibnamefont {Kemp}},
  \bibinfo {author} {\bibfnamefont {L.}~\bibnamefont {Divol}}, \bibinfo
  {author} {\bibfnamefont {D.}~\bibnamefont {Mariscal}}, \bibinfo {author}
  {\bibfnamefont {B.}~\bibnamefont {Westover}}, \bibinfo {author}
  {\bibfnamefont {C.}~\bibnamefont {McGuffey}}, \bibinfo {author}
  {\bibfnamefont {F.~N.}\ \bibnamefont {Beg}}, \bibinfo {author} {\bibfnamefont
  {M.}~\bibnamefont {Suggit}}, \bibinfo {author} {\bibfnamefont
  {C.}~\bibnamefont {Chen}}, \bibinfo {author} {\bibfnamefont {D.}~\bibnamefont
  {Hey}}, \bibinfo {author} {\bibfnamefont {B.}~\bibnamefont {Maddox}},
  \bibinfo {author} {\bibfnamefont {J.}~\bibnamefont {Hawreliak}}, \bibinfo
  {author} {\bibfnamefont {H.-S.}\ \bibnamefont {Park}}, \bibinfo {author}
  {\bibfnamefont {B.}~\bibnamefont {Remington}}, \bibinfo {author}
  {\bibfnamefont {M.~S.}\ \bibnamefont {Wei}},\ and\ \bibinfo {author}
  {\bibfnamefont {A.}~\bibnamefont {MacPhee}},\ }\href
  {https://doi.org/10.1063/1.4865230} {\bibfield  {journal} {\bibinfo
  {journal} {Phys. Plasmas}\ }\textbf {\bibinfo {volume} {21}},\ \bibinfo
  {pages} {031211} (\bibinfo {year} {2014})}\BibitemShut {NoStop}%
\bibitem [{\citenamefont {Cheng}\ \emph {et~al.}(2015)\citenamefont {Cheng},
  \citenamefont {Hill}, \citenamefont {Heubel},\ and\ \citenamefont
  {Vel\'{a}squez-Garc\'{i}a}}]{Cheng2015}%
  \BibitemOpen
  \bibfield  {author} {\bibinfo {author} {\bibfnamefont {S.}~\bibnamefont
  {Cheng}}, \bibinfo {author} {\bibfnamefont {F.~A.}\ \bibnamefont {Hill}},
  \bibinfo {author} {\bibfnamefont {E.~V.}\ \bibnamefont {Heubel}},\ and\
  \bibinfo {author} {\bibfnamefont {L.~F.}\ \bibnamefont
  {Vel\'{a}squez-Garc\'{i}a}},\ }\href
  {https://doi.org/10.1109/JMEMS.2014.2332176} {\bibfield  {journal} {\bibinfo
  {journal} {J. Microelectromechanical Syst.}\ }\textbf {\bibinfo {volume}
  {24}},\ \bibinfo {pages} {373} (\bibinfo {year} {2015})}\BibitemShut
  {NoStop}%
\bibitem [{\citenamefont {Huntington}\ \emph {et~al.}(2018)\citenamefont
  {Huntington}, \citenamefont {McNaney}, \citenamefont {Gumbrell},
  \citenamefont {Krygier}, \citenamefont {Wehrenberg},\ and\ \citenamefont
  {Park}}]{Huntington2018}%
  \BibitemOpen
  \bibfield  {author} {\bibinfo {author} {\bibfnamefont {C.~M.}\ \bibnamefont
  {Huntington}}, \bibinfo {author} {\bibfnamefont {J.~M.}\ \bibnamefont
  {McNaney}}, \bibinfo {author} {\bibfnamefont {E.}~\bibnamefont {Gumbrell}},
  \bibinfo {author} {\bibfnamefont {A.}~\bibnamefont {Krygier}}, \bibinfo
  {author} {\bibfnamefont {C.}~\bibnamefont {Wehrenberg}},\ and\ \bibinfo
  {author} {\bibfnamefont {H.-S.}\ \bibnamefont {Park}},\ }\href
  {https://doi.org/10.1063/1.5039379} {\bibfield  {journal} {\bibinfo
  {journal} {Rev. Sci. Instrum.}\ }\textbf {\bibinfo {volume} {89}},\ \bibinfo
  {pages} {10G121} (\bibinfo {year} {2018})}\BibitemShut {NoStop}%
\bibitem [{\citenamefont {Underwood}\ \emph {et~al.}(2020)\citenamefont
  {Underwood}, \citenamefont {Baird}, \citenamefont {Murphy}, \citenamefont
  {Armstrong}, \citenamefont {Thornton}, \citenamefont {Finlay}, \citenamefont
  {Streeter}, \citenamefont {Selwood}, \citenamefont {Brierley}, \citenamefont
  {Cipiccia}, \citenamefont {Gruse}, \citenamefont {McKenna}, \citenamefont
  {Najmudin}, \citenamefont {Neely}, \citenamefont {Rusby}, \citenamefont
  {Symes},\ and\ \citenamefont {Brenner}}]{Underwood2020}%
  \BibitemOpen
  \bibfield  {author} {\bibinfo {author} {\bibfnamefont {C.~I.~D.}\
  \bibnamefont {Underwood}}, \bibinfo {author} {\bibfnamefont {C.~D.}\
  \bibnamefont {Baird}}, \bibinfo {author} {\bibfnamefont {C.~D.}\ \bibnamefont
  {Murphy}}, \bibinfo {author} {\bibfnamefont {C.~D.}\ \bibnamefont
  {Armstrong}}, \bibinfo {author} {\bibfnamefont {C.}~\bibnamefont {Thornton}},
  \bibinfo {author} {\bibfnamefont {O.~J.}\ \bibnamefont {Finlay}}, \bibinfo
  {author} {\bibfnamefont {M.~J.~V.}\ \bibnamefont {Streeter}}, \bibinfo
  {author} {\bibfnamefont {M.~P.}\ \bibnamefont {Selwood}}, \bibinfo {author}
  {\bibfnamefont {N.}~\bibnamefont {Brierley}}, \bibinfo {author}
  {\bibfnamefont {S.}~\bibnamefont {Cipiccia}}, \bibinfo {author}
  {\bibfnamefont {J.-N.}\ \bibnamefont {Gruse}}, \bibinfo {author}
  {\bibfnamefont {P.}~\bibnamefont {McKenna}}, \bibinfo {author} {\bibfnamefont
  {Z.}~\bibnamefont {Najmudin}}, \bibinfo {author} {\bibfnamefont
  {D.}~\bibnamefont {Neely}}, \bibinfo {author} {\bibfnamefont
  {D.}~\bibnamefont {Rusby}}, \bibinfo {author} {\bibfnamefont {D.~R.}\
  \bibnamefont {Symes}},\ and\ \bibinfo {author} {\bibfnamefont {C.~M.}\
  \bibnamefont {Brenner}},\ }\href {https://doi.org/10.1088/1361-6587/abbebe}
  {\bibfield  {journal} {\bibinfo  {journal} {Plasma Phys. Control. Fusion}\
  }\textbf {\bibinfo {volume} {62}},\ \bibinfo {pages} {124002} (\bibinfo
  {year} {2020})}\BibitemShut {NoStop}%
\bibitem [{\citenamefont {Singh}\ \emph {et~al.}(2021)\citenamefont {Singh},
  \citenamefont {Armstrong}, \citenamefont {Kang}, \citenamefont {Ren},
  \citenamefont {Liu}, \citenamefont {Hua}, \citenamefont {Rusby},
  \citenamefont {Klimo}, \citenamefont {Versaci}, \citenamefont {Zhang},
  \citenamefont {Sun}, \citenamefont {Zhu}, \citenamefont {Lei}, \citenamefont
  {Ouyang}, \citenamefont {Lancia}, \citenamefont {Garcia}, \citenamefont
  {Wagner}, \citenamefont {Cowan}, \citenamefont {Zhu}, \citenamefont
  {Schlegel}, \citenamefont {Weber}, \citenamefont {McKenna}, \citenamefont
  {Neely}, \citenamefont {Tikhonchuk},\ and\ \citenamefont
  {Kumar}}]{Singh2021}%
  \BibitemOpen
  \bibfield  {author} {\bibinfo {author} {\bibfnamefont {S.}~\bibnamefont
  {Singh}}, \bibinfo {author} {\bibfnamefont {C.~D.}\ \bibnamefont
  {Armstrong}}, \bibinfo {author} {\bibfnamefont {N.}~\bibnamefont {Kang}},
  \bibinfo {author} {\bibfnamefont {L.}~\bibnamefont {Ren}}, \bibinfo {author}
  {\bibfnamefont {H.}~\bibnamefont {Liu}}, \bibinfo {author} {\bibfnamefont
  {N.}~\bibnamefont {Hua}}, \bibinfo {author} {\bibfnamefont {D.~R.}\
  \bibnamefont {Rusby}}, \bibinfo {author} {\bibfnamefont {O.}~\bibnamefont
  {Klimo}}, \bibinfo {author} {\bibfnamefont {R.}~\bibnamefont {Versaci}},
  \bibinfo {author} {\bibfnamefont {Y.}~\bibnamefont {Zhang}}, \bibinfo
  {author} {\bibfnamefont {M.}~\bibnamefont {Sun}}, \bibinfo {author}
  {\bibfnamefont {B.}~\bibnamefont {Zhu}}, \bibinfo {author} {\bibfnamefont
  {A.}~\bibnamefont {Lei}}, \bibinfo {author} {\bibfnamefont {X.}~\bibnamefont
  {Ouyang}}, \bibinfo {author} {\bibfnamefont {L.}~\bibnamefont {Lancia}},
  \bibinfo {author} {\bibfnamefont {A.~L.}\ \bibnamefont {Garcia}}, \bibinfo
  {author} {\bibfnamefont {A.}~\bibnamefont {Wagner}}, \bibinfo {author}
  {\bibfnamefont {T.}~\bibnamefont {Cowan}}, \bibinfo {author} {\bibfnamefont
  {J.}~\bibnamefont {Zhu}}, \bibinfo {author} {\bibfnamefont {T.}~\bibnamefont
  {Schlegel}}, \bibinfo {author} {\bibfnamefont {S.}~\bibnamefont {Weber}},
  \bibinfo {author} {\bibfnamefont {P.}~\bibnamefont {McKenna}}, \bibinfo
  {author} {\bibfnamefont {D.}~\bibnamefont {Neely}}, \bibinfo {author}
  {\bibfnamefont {V.}~\bibnamefont {Tikhonchuk}},\ and\ \bibinfo {author}
  {\bibfnamefont {D.}~\bibnamefont {Kumar}},\ }\href
  {https://doi.org/10.1088/1361-6587/abcf7e} {\bibfield  {journal} {\bibinfo
  {journal} {Plasma Phys. Control. Fusion}\ }\textbf {\bibinfo {volume} {63}},\
  \bibinfo {pages} {035004} (\bibinfo {year} {2021})}\BibitemShut {NoStop}%
\bibitem [{\citenamefont {Vysko\v{c}il}\ \emph {et~al.}(2018)\citenamefont
  {Vysko\v{c}il}, \citenamefont {Klimo},\ and\ \citenamefont
  {Weber}}]{Vyskocil2018}%
  \BibitemOpen
  \bibfield  {author} {\bibinfo {author} {\bibfnamefont {J.}~\bibnamefont
  {Vysko\v{c}il}}, \bibinfo {author} {\bibfnamefont {O.}~\bibnamefont
  {Klimo}},\ and\ \bibinfo {author} {\bibfnamefont {S.}~\bibnamefont {Weber}},\
  }\href {https://doi.org/10.1088/1361-6587/aab4c3} {\bibfield  {journal}
  {\bibinfo  {journal} {Plasma Phys. Control. Fusion}\ }\textbf {\bibinfo
  {volume} {60}},\ \bibinfo {pages} {054013} (\bibinfo {year}
  {2018})}\BibitemShut {NoStop}%
\bibitem [{\citenamefont {Bakhtiari}\ \emph {et~al.}(2005)\citenamefont
  {Bakhtiari}, \citenamefont {Kramer}, \citenamefont {Takechi}, \citenamefont
  {Tamai}, \citenamefont {Miura}, \citenamefont {Kusama},\ and\ \citenamefont
  {Kamada}}]{Bakhtiari2005}%
  \BibitemOpen
  \bibfield  {author} {\bibinfo {author} {\bibfnamefont {M.}~\bibnamefont
  {Bakhtiari}}, \bibinfo {author} {\bibfnamefont {G.~J.}\ \bibnamefont
  {Kramer}}, \bibinfo {author} {\bibfnamefont {M.}~\bibnamefont {Takechi}},
  \bibinfo {author} {\bibfnamefont {H.}~\bibnamefont {Tamai}}, \bibinfo
  {author} {\bibfnamefont {Y.}~\bibnamefont {Miura}}, \bibinfo {author}
  {\bibfnamefont {Y.}~\bibnamefont {Kusama}},\ and\ \bibinfo {author}
  {\bibfnamefont {Y.}~\bibnamefont {Kamada}},\ }\href
  {https://doi.org/10.1103/PhysRevLett.94.215003} {\bibfield  {journal}
  {\bibinfo  {journal} {Phys. Rev. Lett.}\ }\textbf {\bibinfo {volume} {94}},\
  \bibinfo {pages} {215003} (\bibinfo {year} {2005})}\BibitemShut {NoStop}%
\bibitem [{\citenamefont {Embr\'{e}us}\ \emph {et~al.}(2016)\citenamefont
  {Embr\'{e}us}, \citenamefont {Stahl},\ and\ \citenamefont
  {F\"{u}l\"{o}p}}]{Embreus2016}%
  \BibitemOpen
  \bibfield  {author} {\bibinfo {author} {\bibfnamefont {O.}~\bibnamefont
  {Embr\'{e}us}}, \bibinfo {author} {\bibfnamefont {A.}~\bibnamefont {Stahl}},\
  and\ \bibinfo {author} {\bibfnamefont {T.}~\bibnamefont {F\"{u}l\"{o}p}},\
  }\href {https://doi.org/10.1088/1367-2630/18/9/093023} {\bibfield  {journal}
  {\bibinfo  {journal} {New J. Phys.}\ }\textbf {\bibinfo {volume} {18}},\
  \bibinfo {pages} {093023} (\bibinfo {year} {2016})}\BibitemShut {NoStop}%
\bibitem [{\citenamefont {{Kellogg}}\ \emph {et~al.}(1975)\citenamefont
  {{Kellogg}}, \citenamefont {{Baldwin}},\ and\ \citenamefont
  {{Koch}}}]{Kellogg1975}%
  \BibitemOpen
  \bibfield  {author} {\bibinfo {author} {\bibfnamefont {E.}~\bibnamefont
  {{Kellogg}}}, \bibinfo {author} {\bibfnamefont {J.~R.}\ \bibnamefont
  {{Baldwin}}},\ and\ \bibinfo {author} {\bibfnamefont {D.}~\bibnamefont
  {{Koch}}},\ }\href {https://doi.org/10.1086/153692} {\bibfield  {journal}
  {\bibinfo  {journal} {Astrophys. J.}\ }\textbf {\bibinfo {volume} {199}},\
  \bibinfo {pages} {299} (\bibinfo {year} {1975})}\BibitemShut {NoStop}%
\bibitem [{\citenamefont {{Svensson}}(1982)}]{Svensson1982}%
  \BibitemOpen
  \bibfield  {author} {\bibinfo {author} {\bibfnamefont {R.}~\bibnamefont
  {{Svensson}}},\ }\href {https://doi.org/10.1086/160082} {\bibfield  {journal}
  {\bibinfo  {journal} {Astrophys. J.}\ }\textbf {\bibinfo {volume} {258}},\
  \bibinfo {pages} {335} (\bibinfo {year} {1982})}\BibitemShut {NoStop}%
\bibitem [{\citenamefont {Chluba}\ \emph {et~al.}(2020)\citenamefont {Chluba},
  \citenamefont {Ravenni},\ and\ \citenamefont {Bolliet}}]{Chluba2020}%
  \BibitemOpen
  \bibfield  {author} {\bibinfo {author} {\bibfnamefont {J.}~\bibnamefont
  {Chluba}}, \bibinfo {author} {\bibfnamefont {A.}~\bibnamefont {Ravenni}},\
  and\ \bibinfo {author} {\bibfnamefont {B.}~\bibnamefont {Bolliet}},\ }\href
  {https://doi.org/10.1093/mnras/stz3389} {\bibfield  {journal} {\bibinfo
  {journal} {Mon. Not. R. Astron. Soc.}\ }\textbf {\bibinfo {volume} {492}},\
  \bibinfo {pages} {177} (\bibinfo {year} {2020})}\BibitemShut {NoStop}%
\bibitem [{\citenamefont {Pradler}\ and\ \citenamefont
  {Semmelrock}(2021{\natexlab{a}})}]{Pradler2021a}%
  \BibitemOpen
  \bibfield  {author} {\bibinfo {author} {\bibfnamefont {J.}~\bibnamefont
  {Pradler}}\ and\ \bibinfo {author} {\bibfnamefont {L.}~\bibnamefont
  {Semmelrock}},\ }\href {https://doi.org/10.3847/1538-4357/abc9c5} {\bibfield
  {journal} {\bibinfo  {journal} {Astrophys. J.}\ }\textbf {\bibinfo {volume}
  {909}},\ \bibinfo {pages} {134} (\bibinfo {year}
  {2021}{\natexlab{a}})}\BibitemShut {NoStop}%
\bibitem [{\citenamefont {Pradler}\ and\ \citenamefont
  {Semmelrock}(2021{\natexlab{b}})}]{Pradler2021b}%
  \BibitemOpen
  \bibfield  {author} {\bibinfo {author} {\bibfnamefont {J.}~\bibnamefont
  {Pradler}}\ and\ \bibinfo {author} {\bibfnamefont {L.}~\bibnamefont
  {Semmelrock}},\ }\href {https://doi.org/10.3847/1538-4357/ac0898} {\bibfield
  {journal} {\bibinfo  {journal} {Astrophys. J.}\ }\textbf {\bibinfo {volume}
  {916}},\ \bibinfo {pages} {105} (\bibinfo {year}
  {2021}{\natexlab{b}})}\BibitemShut {NoStop}%
\bibitem [{\citenamefont {Pradler}\ and\ \citenamefont
  {Semmelrock}(2021{\natexlab{c}})}]{Pradler2021c}%
  \BibitemOpen
  \bibfield  {author} {\bibinfo {author} {\bibfnamefont {J.}~\bibnamefont
  {Pradler}}\ and\ \bibinfo {author} {\bibfnamefont {L.}~\bibnamefont
  {Semmelrock}},\ }\href {https://doi.org/10.3847/1538-4357/ac24a8} {\bibfield
  {journal} {\bibinfo  {journal} {Astrophys. J.}\ }\textbf {\bibinfo {volume}
  {922}},\ \bibinfo {pages} {57} (\bibinfo {year}
  {2021}{\natexlab{c}})}\BibitemShut {NoStop}%
\bibitem [{\citenamefont {Sarazin}(1986)}]{Sarazin1986}%
  \BibitemOpen
  \bibfield  {author} {\bibinfo {author} {\bibfnamefont {C.~L.}\ \bibnamefont
  {Sarazin}},\ }\href {https://doi.org/10.1103/RevModPhys.58.1} {\bibfield
  {journal} {\bibinfo  {journal} {Rev. Mod. Phys.}\ }\textbf {\bibinfo {volume}
  {58}},\ \bibinfo {pages} {1} (\bibinfo {year} {1986})}\BibitemShut {NoStop}%
\bibitem [{\citenamefont {Sarazin}\ and\ \citenamefont
  {Kempner}(2000)}]{Sarazin2000}%
  \BibitemOpen
  \bibfield  {author} {\bibinfo {author} {\bibfnamefont {C.~L.}\ \bibnamefont
  {Sarazin}}\ and\ \bibinfo {author} {\bibfnamefont {J.~C.}\ \bibnamefont
  {Kempner}},\ }\href {https://doi.org/10.1086/308649} {\bibfield  {journal}
  {\bibinfo  {journal} {Astrophys. J.}\ }\textbf {\bibinfo {volume} {533}},\
  \bibinfo {pages} {73} (\bibinfo {year} {2000})}\BibitemShut {NoStop}%
\bibitem [{\citenamefont {Pearce}\ \emph {et~al.}(2000)\citenamefont {Pearce},
  \citenamefont {Thomas}, \citenamefont {Couchman},\ and\ \citenamefont
  {Edge}}]{Pearce2000}%
  \BibitemOpen
  \bibfield  {author} {\bibinfo {author} {\bibfnamefont {F.~R.}\ \bibnamefont
  {Pearce}}, \bibinfo {author} {\bibfnamefont {P.~A.}\ \bibnamefont {Thomas}},
  \bibinfo {author} {\bibfnamefont {H.~M.~P.}\ \bibnamefont {Couchman}},\ and\
  \bibinfo {author} {\bibfnamefont {A.~C.}\ \bibnamefont {Edge}},\ }\href
  {https://doi.org/10.1046/j.1365-8711.2000.03773.x} {\bibfield  {journal}
  {\bibinfo  {journal} {Mon. Not. R. Astron. Soc.}\ }\textbf {\bibinfo {volume}
  {317}},\ \bibinfo {pages} {1029} (\bibinfo {year} {2000})}\BibitemShut
  {NoStop}%
\bibitem [{\citenamefont {Haug}(1975{\natexlab{c}})}]{Haug1975c}%
  \BibitemOpen
  \bibfield  {author} {\bibinfo {author} {\bibfnamefont {E.}~\bibnamefont
  {Haug}},\ }\href {https://doi.org/10.1007/BF00158461} {\bibfield  {journal}
  {\bibinfo  {journal} {Sol. Phys.}\ }\textbf {\bibinfo {volume} {45}},\
  \bibinfo {pages} {453} (\bibinfo {year} {1975}{\natexlab{c}})}\BibitemShut
  {NoStop}%
\bibitem [{\citenamefont {Holman}\ \emph {et~al.}(2003)\citenamefont {Holman},
  \citenamefont {Sui}, \citenamefont {Schwartz},\ and\ \citenamefont
  {Emslie}}]{Holman2003}%
  \BibitemOpen
  \bibfield  {author} {\bibinfo {author} {\bibfnamefont {G.~D.}\ \bibnamefont
  {Holman}}, \bibinfo {author} {\bibfnamefont {L.}~\bibnamefont {Sui}},
  \bibinfo {author} {\bibfnamefont {R.~A.}\ \bibnamefont {Schwartz}},\ and\
  \bibinfo {author} {\bibfnamefont {A.~G.}\ \bibnamefont {Emslie}},\ }\href
  {https://doi.org/10.1086/378488} {\bibfield  {journal} {\bibinfo  {journal}
  {Astrophys. J.}\ }\textbf {\bibinfo {volume} {595}},\ \bibinfo {pages} {L97}
  (\bibinfo {year} {2003})}\BibitemShut {NoStop}%
\bibitem [{\citenamefont {Kontar}\ \emph {et~al.}(2007)\citenamefont {Kontar},
  \citenamefont {Emslie}, \citenamefont {Massone}, \citenamefont {Piana},
  \citenamefont {Brown},\ and\ \citenamefont {Prato}}]{Kontar2007}%
  \BibitemOpen
  \bibfield  {author} {\bibinfo {author} {\bibfnamefont {E.~P.}\ \bibnamefont
  {Kontar}}, \bibinfo {author} {\bibfnamefont {A.~G.}\ \bibnamefont {Emslie}},
  \bibinfo {author} {\bibfnamefont {A.~M.}\ \bibnamefont {Massone}}, \bibinfo
  {author} {\bibfnamefont {M.}~\bibnamefont {Piana}}, \bibinfo {author}
  {\bibfnamefont {J.~C.}\ \bibnamefont {Brown}},\ and\ \bibinfo {author}
  {\bibfnamefont {M.}~\bibnamefont {Prato}},\ }\href
  {https://doi.org/10.1086/521977} {\bibfield  {journal} {\bibinfo  {journal}
  {Astrophys. J.}\ }\textbf {\bibinfo {volume} {670}},\ \bibinfo {pages} {857}
  (\bibinfo {year} {2007})}\BibitemShut {NoStop}%
\bibitem [{\citenamefont {Sunyaev}\ and\ \citenamefont
  {Zeldovich}(1970)}]{Sunyaev1970}%
  \BibitemOpen
  \bibfield  {author} {\bibinfo {author} {\bibfnamefont {R.~A.}\ \bibnamefont
  {Sunyaev}}\ and\ \bibinfo {author} {\bibfnamefont {Y.~B.}\ \bibnamefont
  {Zeldovich}},\ }\href {https://doi.org/10.1007/BF00653472} {\bibfield
  {journal} {\bibinfo  {journal} {Astrophys. Space Sci.}\ }\textbf {\bibinfo
  {volume} {7}},\ \bibinfo {pages} {20} (\bibinfo {year} {1970})}\BibitemShut
  {NoStop}%
\bibitem [{\citenamefont {Chluba}\ and\ \citenamefont
  {Sunyaev}(2012)}]{Chluba2012}%
  \BibitemOpen
  \bibfield  {author} {\bibinfo {author} {\bibfnamefont {J.}~\bibnamefont
  {Chluba}}\ and\ \bibinfo {author} {\bibfnamefont {R.~A.}\ \bibnamefont
  {Sunyaev}},\ }\href {https://doi.org/10.1111/j.1365-2966.2011.19786.x}
  {\bibfield  {journal} {\bibinfo  {journal} {Mon. Not. R. Astron. Soc.}\
  }\textbf {\bibinfo {volume} {419}},\ \bibinfo {pages} {1294} (\bibinfo {year}
  {2012})}\BibitemShut {NoStop}%
\bibitem [{\citenamefont {{Narayan}}\ and\ \citenamefont
  {{Yi}}(1995)}]{Narayan1994}%
  \BibitemOpen
  \bibfield  {author} {\bibinfo {author} {\bibfnamefont {R.}~\bibnamefont
  {{Narayan}}}\ and\ \bibinfo {author} {\bibfnamefont {I.}~\bibnamefont
  {{Yi}}},\ }\href {https://doi.org/10.1086/176343} {\bibfield  {journal}
  {\bibinfo  {journal} {Astrophys. J.}\ }\textbf {\bibinfo {volume} {452}},\
  \bibinfo {pages} {710} (\bibinfo {year} {1995})}\BibitemShut {NoStop}%
\bibitem [{\citenamefont {Yarza}\ \emph {et~al.}(2020)\citenamefont {Yarza},
  \citenamefont {Wong}, \citenamefont {Ryan},\ and\ \citenamefont
  {Gammie}}]{Yarza2020}%
  \BibitemOpen
  \bibfield  {author} {\bibinfo {author} {\bibfnamefont {R.}~\bibnamefont
  {Yarza}}, \bibinfo {author} {\bibfnamefont {G.~N.}\ \bibnamefont {Wong}},
  \bibinfo {author} {\bibfnamefont {B.~R.}\ \bibnamefont {Ryan}},\ and\
  \bibinfo {author} {\bibfnamefont {C.~F.}\ \bibnamefont {Gammie}},\ }\href
  {https://doi.org/10.3847/1538-4357/ab9808} {\bibfield  {journal} {\bibinfo
  {journal} {Astrophys. J.}\ }\textbf {\bibinfo {volume} {898}},\ \bibinfo
  {pages} {50} (\bibinfo {year} {2020})}\BibitemShut {NoStop}%
\bibitem [{\citenamefont {Keefe}(1982)}]{Keefe1982}%
  \BibitemOpen
  \bibfield  {author} {\bibinfo {author} {\bibfnamefont {D.}~\bibnamefont
  {Keefe}},\ }\href {https://doi.org/10.1146/annurev.ns.32.120182.002135}
  {\bibfield  {journal} {\bibinfo  {journal} {Annu. Rev. Nucl. Part. Sci.}\
  }\textbf {\bibinfo {volume} {32}},\ \bibinfo {pages} {391} (\bibinfo {year}
  {1982})}\BibitemShut {NoStop}%
\bibitem [{\citenamefont {Craxton}\ \emph {et~al.}(2015)\citenamefont
  {Craxton}, \citenamefont {Anderson}, \citenamefont {Boehly}, \citenamefont
  {Goncharov}, \citenamefont {Harding}, \citenamefont {Knauer}, \citenamefont
  {McCrory}, \citenamefont {McKenty}, \citenamefont {Meyerhofer}, \citenamefont
  {Myatt}, \citenamefont {Schmitt}, \citenamefont {Sethian}, \citenamefont
  {Short}, \citenamefont {Skupsky}, \citenamefont {Theobald}, \citenamefont
  {Kruer}, \citenamefont {Tanaka}, \citenamefont {Betti}, \citenamefont
  {Collins}, \citenamefont {Delettrez}, \citenamefont {Hu}, \citenamefont
  {Marozas}, \citenamefont {Maximov}, \citenamefont {Michel}, \citenamefont
  {Radha}, \citenamefont {Regan}, \citenamefont {Sangster}, \citenamefont
  {Seka}, \citenamefont {Solodov}, \citenamefont {Soures}, \citenamefont
  {Stoeckl},\ and\ \citenamefont {Zuegel}}]{Craxton2015}%
  \BibitemOpen
  \bibfield  {author} {\bibinfo {author} {\bibfnamefont {R.~S.}\ \bibnamefont
  {Craxton}}, \bibinfo {author} {\bibfnamefont {K.~S.}\ \bibnamefont
  {Anderson}}, \bibinfo {author} {\bibfnamefont {T.~R.}\ \bibnamefont
  {Boehly}}, \bibinfo {author} {\bibfnamefont {V.~N.}\ \bibnamefont
  {Goncharov}}, \bibinfo {author} {\bibfnamefont {D.~R.}\ \bibnamefont
  {Harding}}, \bibinfo {author} {\bibfnamefont {J.~P.}\ \bibnamefont {Knauer}},
  \bibinfo {author} {\bibfnamefont {R.~L.}\ \bibnamefont {McCrory}}, \bibinfo
  {author} {\bibfnamefont {P.~W.}\ \bibnamefont {McKenty}}, \bibinfo {author}
  {\bibfnamefont {D.~D.}\ \bibnamefont {Meyerhofer}}, \bibinfo {author}
  {\bibfnamefont {J.~F.}\ \bibnamefont {Myatt}}, \bibinfo {author}
  {\bibfnamefont {A.~J.}\ \bibnamefont {Schmitt}}, \bibinfo {author}
  {\bibfnamefont {J.~D.}\ \bibnamefont {Sethian}}, \bibinfo {author}
  {\bibfnamefont {R.~W.}\ \bibnamefont {Short}}, \bibinfo {author}
  {\bibfnamefont {S.}~\bibnamefont {Skupsky}}, \bibinfo {author} {\bibfnamefont
  {W.}~\bibnamefont {Theobald}}, \bibinfo {author} {\bibfnamefont {W.~L.}\
  \bibnamefont {Kruer}}, \bibinfo {author} {\bibfnamefont {K.}~\bibnamefont
  {Tanaka}}, \bibinfo {author} {\bibfnamefont {R.}~\bibnamefont {Betti}},
  \bibinfo {author} {\bibfnamefont {T.~J.~B.}\ \bibnamefont {Collins}},
  \bibinfo {author} {\bibfnamefont {J.~A.}\ \bibnamefont {Delettrez}}, \bibinfo
  {author} {\bibfnamefont {S.~X.}\ \bibnamefont {Hu}}, \bibinfo {author}
  {\bibfnamefont {J.~A.}\ \bibnamefont {Marozas}}, \bibinfo {author}
  {\bibfnamefont {A.~V.}\ \bibnamefont {Maximov}}, \bibinfo {author}
  {\bibfnamefont {D.~T.}\ \bibnamefont {Michel}}, \bibinfo {author}
  {\bibfnamefont {P.~B.}\ \bibnamefont {Radha}}, \bibinfo {author}
  {\bibfnamefont {S.~P.}\ \bibnamefont {Regan}}, \bibinfo {author}
  {\bibfnamefont {T.~C.}\ \bibnamefont {Sangster}}, \bibinfo {author}
  {\bibfnamefont {W.}~\bibnamefont {Seka}}, \bibinfo {author} {\bibfnamefont
  {A.~A.}\ \bibnamefont {Solodov}}, \bibinfo {author} {\bibfnamefont {J.~M.}\
  \bibnamefont {Soures}}, \bibinfo {author} {\bibfnamefont {C.}~\bibnamefont
  {Stoeckl}},\ and\ \bibinfo {author} {\bibfnamefont {J.~D.}\ \bibnamefont
  {Zuegel}},\ }\href {https://doi.org/10.1063/1.4934714} {\bibfield  {journal}
  {\bibinfo  {journal} {Phys. Plasmas}\ }\textbf {\bibinfo {volume} {22}},\
  \bibinfo {pages} {110501} (\bibinfo {year} {2015})}\BibitemShut {NoStop}%
\bibitem [{\citenamefont {Firouzi~Farrashbandi}\ and\ \citenamefont
  {Eslami-Kalantari}(2020)}]{FirouziFarrashbandi2020}%
  \BibitemOpen
  \bibfield  {author} {\bibinfo {author} {\bibfnamefont {N.}~\bibnamefont
  {Firouzi~Farrashbandi}}\ and\ \bibinfo {author} {\bibfnamefont
  {M.}~\bibnamefont {Eslami-Kalantari}},\ }\href
  {https://doi.org/10.1007/s40094-020-00375-4} {\bibfield  {journal} {\bibinfo
  {journal} {J. Theor. Appl. Phys.}\ }\textbf {\bibinfo {volume} {14}},\
  \bibinfo {pages} {261} (\bibinfo {year} {2020})}\BibitemShut {NoStop}%
\bibitem [{\citenamefont {Rand}(1964)}]{Rand1964}%
  \BibitemOpen
  \bibfield  {author} {\bibinfo {author} {\bibfnamefont {S.}~\bibnamefont
  {Rand}},\ }\href {https://doi.org/10.1103/PhysRev.136.B231} {\bibfield
  {journal} {\bibinfo  {journal} {Phys. Rev.}\ }\textbf {\bibinfo {volume}
  {136}},\ \bibinfo {pages} {B231} (\bibinfo {year} {1964})}\BibitemShut
  {NoStop}%
\bibitem [{\citenamefont {Shima}\ and\ \citenamefont
  {Yatom}(1975)}]{Shima1975}%
  \BibitemOpen
  \bibfield  {author} {\bibinfo {author} {\bibfnamefont {Y.}~\bibnamefont
  {Shima}}\ and\ \bibinfo {author} {\bibfnamefont {H.}~\bibnamefont {Yatom}},\
  }\href {https://doi.org/10.1103/PhysRevA.12.2106} {\bibfield  {journal}
  {\bibinfo  {journal} {Phys. Rev. A}\ }\textbf {\bibinfo {volume} {12}},\
  \bibinfo {pages} {2106} (\bibinfo {year} {1975})}\BibitemShut {NoStop}%
\bibitem [{\citenamefont {Brysk}(1975)}]{Brysk1975}%
  \BibitemOpen
  \bibfield  {author} {\bibinfo {author} {\bibfnamefont {H.}~\bibnamefont
  {Brysk}},\ }\href {https://doi.org/10.1088/0305-4470/8/8/011} {\bibfield
  {journal} {\bibinfo  {journal} {J. Phys. A}\ }\textbf {\bibinfo {volume}
  {8}},\ \bibinfo {pages} {1260} (\bibinfo {year} {1975})}\BibitemShut
  {NoStop}%
\bibitem [{\citenamefont {Friedland}(1979)}]{Friedland1979}%
  \BibitemOpen
  \bibfield  {author} {\bibinfo {author} {\bibfnamefont {L.}~\bibnamefont
  {Friedland}},\ }\href {https://doi.org/10.1088/0022-3700/12/3/018} {\bibfield
   {journal} {\bibinfo  {journal} {J. Phys. B}\ }\textbf {\bibinfo {volume}
  {12}},\ \bibinfo {pages} {409} (\bibinfo {year} {1979})}\BibitemShut
  {NoStop}%
\bibitem [{\citenamefont {Mora}(1982)}]{Mora1982}%
  \BibitemOpen
  \bibfield  {author} {\bibinfo {author} {\bibfnamefont {P.}~\bibnamefont
  {Mora}},\ }\href {https://doi.org/10.1063/1.863837} {\bibfield  {journal}
  {\bibinfo  {journal} {Phys. Fluids}\ }\textbf {\bibinfo {volume} {25}},\
  \bibinfo {pages} {1051} (\bibinfo {year} {1982})}\BibitemShut {NoStop}%
\bibitem [{\citenamefont {Balescu}(1982)}]{Balescu1982}%
  \BibitemOpen
  \bibfield  {author} {\bibinfo {author} {\bibfnamefont {R.}~\bibnamefont
  {Balescu}},\ }\href {https://doi.org/10.1017/S0022377800011089} {\bibfield
  {journal} {\bibinfo  {journal} {J. Plasma Phys.}\ }\textbf {\bibinfo {volume}
  {27}},\ \bibinfo {pages} {553} (\bibinfo {year} {1982})}\BibitemShut
  {NoStop}%
\bibitem [{\citenamefont {Garban-Labaune}\ \emph {et~al.}(1982)\citenamefont
  {Garban-Labaune}, \citenamefont {Fabre}, \citenamefont {Max}, \citenamefont
  {Fabbro}, \citenamefont {Amiranoff}, \citenamefont {Virmont}, \citenamefont
  {Weinfeld},\ and\ \citenamefont {Michard}}]{GarbanLabaune1982}%
  \BibitemOpen
  \bibfield  {author} {\bibinfo {author} {\bibfnamefont {C.}~\bibnamefont
  {Garban-Labaune}}, \bibinfo {author} {\bibfnamefont {E.}~\bibnamefont
  {Fabre}}, \bibinfo {author} {\bibfnamefont {C.~E.}\ \bibnamefont {Max}},
  \bibinfo {author} {\bibfnamefont {R.}~\bibnamefont {Fabbro}}, \bibinfo
  {author} {\bibfnamefont {F.}~\bibnamefont {Amiranoff}}, \bibinfo {author}
  {\bibfnamefont {J.}~\bibnamefont {Virmont}}, \bibinfo {author} {\bibfnamefont
  {M.}~\bibnamefont {Weinfeld}},\ and\ \bibinfo {author} {\bibfnamefont
  {A.}~\bibnamefont {Michard}},\ }\href
  {https://doi.org/10.1103/PhysRevLett.48.1018} {\bibfield  {journal} {\bibinfo
   {journal} {Phys. Rev. Lett.}\ }\textbf {\bibinfo {volume} {48}},\ \bibinfo
  {pages} {1018} (\bibinfo {year} {1982})}\BibitemShut {NoStop}%
\bibitem [{\citenamefont {Cauble}\ and\ \citenamefont
  {Rozmus}(1985)}]{Cauble1985}%
  \BibitemOpen
  \bibfield  {author} {\bibinfo {author} {\bibfnamefont {R.}~\bibnamefont
  {Cauble}}\ and\ \bibinfo {author} {\bibfnamefont {W.}~\bibnamefont
  {Rozmus}},\ }\href {https://doi.org/10.1063/1.865338} {\bibfield  {journal}
  {\bibinfo  {journal} {Phys. Fluids}\ }\textbf {\bibinfo {volume} {28}},\
  \bibinfo {pages} {3387} (\bibinfo {year} {1985})}\BibitemShut {NoStop}%
\bibitem [{\citenamefont {Skupsky}(1987)}]{Skupsky1987}%
  \BibitemOpen
  \bibfield  {author} {\bibinfo {author} {\bibfnamefont {S.}~\bibnamefont
  {Skupsky}},\ }\href {https://doi.org/10.1103/PhysRevA.36.5701} {\bibfield
  {journal} {\bibinfo  {journal} {Phys. Rev. A}\ }\textbf {\bibinfo {volume}
  {36}},\ \bibinfo {pages} {5701} (\bibinfo {year} {1987})}\BibitemShut
  {NoStop}%
\bibitem [{\citenamefont {Tsytovich}(1995)}]{Tsytovich1995}%
  \BibitemOpen
  \bibfield  {author} {\bibinfo {author} {\bibfnamefont {V.~N.}\ \bibnamefont
  {Tsytovich}},\ }\href {https://doi.org/10.1070/PU1995v038n01ABEH000065}
  {\bibfield  {journal} {\bibinfo  {journal} {Phys. Usp.}\ }\textbf {\bibinfo
  {volume} {38}},\ \bibinfo {pages} {87} (\bibinfo {year} {1995})}\BibitemShut
  {NoStop}%
\bibitem [{\citenamefont {Tsytovich}\ \emph {et~al.}(1996)\citenamefont
  {Tsytovich}, \citenamefont {Bingham}, \citenamefont {de~Angelis},\ and\
  \citenamefont {Forlani}}]{Tsytovich1996}%
  \BibitemOpen
  \bibfield  {author} {\bibinfo {author} {\bibfnamefont {V.~N.}\ \bibnamefont
  {Tsytovich}}, \bibinfo {author} {\bibfnamefont {R.}~\bibnamefont {Bingham}},
  \bibinfo {author} {\bibfnamefont {U.}~\bibnamefont {de~Angelis}},\ and\
  \bibinfo {author} {\bibfnamefont {A.}~\bibnamefont {Forlani}},\ }\href
  {https://doi.org/10.1017/S0022377800019140} {\bibfield  {journal} {\bibinfo
  {journal} {J. Plasma Phys.}\ }\textbf {\bibinfo {volume} {56}},\ \bibinfo
  {pages} {127} (\bibinfo {year} {1996})}\BibitemShut {NoStop}%
\bibitem [{\citenamefont {Tsallis}\ and\ \citenamefont {{de
  Souza}}(1997)}]{Tsallis1997}%
  \BibitemOpen
  \bibfield  {author} {\bibinfo {author} {\bibfnamefont {C.}~\bibnamefont
  {Tsallis}}\ and\ \bibinfo {author} {\bibfnamefont {A.}~\bibnamefont {{de
  Souza}}},\ }\href
  {https://doi.org/https://doi.org/10.1016/S0375-9601(97)00676-2} {\bibfield
  {journal} {\bibinfo  {journal} {Phys. Lett. A}\ }\textbf {\bibinfo {volume}
  {235}},\ \bibinfo {pages} {444} (\bibinfo {year} {1997})}\BibitemShut
  {NoStop}%
\bibitem [{\citenamefont {Kostyukov}(2001)}]{Kostyukov2001}%
  \BibitemOpen
  \bibfield  {author} {\bibinfo {author} {\bibfnamefont {I.~Y.}\ \bibnamefont
  {Kostyukov}},\ }\href {https://doi.org/10.1134/1.1381634} {\bibfield
  {journal} {\bibinfo  {journal} {J. Exp. Theor. Phys. Lett.}\ }\textbf
  {\bibinfo {volume} {73}},\ \bibinfo {pages} {393} (\bibinfo {year}
  {2001})}\BibitemShut {NoStop}%
\bibitem [{\citenamefont {Wierling}\ \emph {et~al.}(2001)\citenamefont
  {Wierling}, \citenamefont {Millat}, \citenamefont {R\"{o}pke}, \citenamefont
  {Redmer},\ and\ \citenamefont {Reinholz}}]{Wierling2001}%
  \BibitemOpen
  \bibfield  {author} {\bibinfo {author} {\bibfnamefont {A.}~\bibnamefont
  {Wierling}}, \bibinfo {author} {\bibfnamefont {T.}~\bibnamefont {Millat}},
  \bibinfo {author} {\bibfnamefont {G.}~\bibnamefont {R\"{o}pke}}, \bibinfo
  {author} {\bibfnamefont {R.}~\bibnamefont {Redmer}},\ and\ \bibinfo {author}
  {\bibfnamefont {H.}~\bibnamefont {Reinholz}},\ }\href
  {https://doi.org/10.1063/1.1383025} {\bibfield  {journal} {\bibinfo
  {journal} {Phys. Plasmas}\ }\textbf {\bibinfo {volume} {8}},\ \bibinfo
  {pages} {3810} (\bibinfo {year} {2001})}\BibitemShut {NoStop}%
\bibitem [{\citenamefont {Berger}\ \emph {et~al.}(2004)\citenamefont {Berger},
  \citenamefont {Clark}, \citenamefont {Solodov}, \citenamefont {Valeo},\ and\
  \citenamefont {Fisch}}]{Berger2004}%
  \BibitemOpen
  \bibfield  {author} {\bibinfo {author} {\bibfnamefont {R.~L.}\ \bibnamefont
  {Berger}}, \bibinfo {author} {\bibfnamefont {D.~S.}\ \bibnamefont {Clark}},
  \bibinfo {author} {\bibfnamefont {A.~A.}\ \bibnamefont {Solodov}}, \bibinfo
  {author} {\bibfnamefont {E.~J.}\ \bibnamefont {Valeo}},\ and\ \bibinfo
  {author} {\bibfnamefont {N.~J.}\ \bibnamefont {Fisch}},\ }\href
  {https://doi.org/10.1063/1.1695356} {\bibfield  {journal} {\bibinfo
  {journal} {Phys. Plasmas}\ }\textbf {\bibinfo {volume} {11}},\ \bibinfo
  {pages} {1931} (\bibinfo {year} {2004})}\BibitemShut {NoStop}%
\bibitem [{\citenamefont {Avetissian}\ \emph {et~al.}(2013)\citenamefont
  {Avetissian}, \citenamefont {Ghazaryan},\ and\ \citenamefont
  {Mkrtchian}}]{Avetissian2013}%
  \BibitemOpen
  \bibfield  {author} {\bibinfo {author} {\bibfnamefont {H.~K.}\ \bibnamefont
  {Avetissian}}, \bibinfo {author} {\bibfnamefont {A.~G.}\ \bibnamefont
  {Ghazaryan}},\ and\ \bibinfo {author} {\bibfnamefont {G.~F.}\ \bibnamefont
  {Mkrtchian}},\ }\href {https://doi.org/10.1088/0953-4075/46/20/205701}
  {\bibfield  {journal} {\bibinfo  {journal} {J. Phys. B}\ }\textbf {\bibinfo
  {volume} {46}},\ \bibinfo {pages} {205701} (\bibinfo {year}
  {2013})}\BibitemShut {NoStop}%
\bibitem [{\citenamefont {{Firouzi Farrashbandi}}\ \emph
  {et~al.}(2015)\citenamefont {{Firouzi Farrashbandi}}, \citenamefont
  {Gholamzadeh}, \citenamefont {Eslami-Kalantari}, \citenamefont {Sharifian},\
  and\ \citenamefont {Sid}}]{FirouziFarrashbandi2015}%
  \BibitemOpen
  \bibfield  {author} {\bibinfo {author} {\bibfnamefont {N.}~\bibnamefont
  {{Firouzi Farrashbandi}}}, \bibinfo {author} {\bibfnamefont {L.}~\bibnamefont
  {Gholamzadeh}}, \bibinfo {author} {\bibfnamefont {M.}~\bibnamefont
  {Eslami-Kalantari}}, \bibinfo {author} {\bibfnamefont {M.}~\bibnamefont
  {Sharifian}},\ and\ \bibinfo {author} {\bibfnamefont {A.}~\bibnamefont
  {Sid}},\ }\href {https://doi.org/https://doi.org/10.1016/j.hedp.2015.05.002}
  {\bibfield  {journal} {\bibinfo  {journal} {High Energy Density Phys.}\
  }\textbf {\bibinfo {volume} {16}},\ \bibinfo {pages} {32} (\bibinfo {year}
  {2015})}\BibitemShut {NoStop}%
\bibitem [{\citenamefont {Farrashbandi}\ \emph {et~al.}(2020)\citenamefont
  {Farrashbandi}, \citenamefont {Eslami-Kalantari},\ and\ \citenamefont
  {Sid}}]{FirouziFarrashbandi2020b}%
  \BibitemOpen
  \bibfield  {author} {\bibinfo {author} {\bibfnamefont {N.~F.}\ \bibnamefont
  {Farrashbandi}}, \bibinfo {author} {\bibfnamefont {M.}~\bibnamefont
  {Eslami-Kalantari}},\ and\ \bibinfo {author} {\bibfnamefont {A.}~\bibnamefont
  {Sid}},\ }\href {https://doi.org/10.1209/0295-5075/130/25001} {\bibfield
  {journal} {\bibinfo  {journal} {EPL}\ }\textbf {\bibinfo {volume} {130}},\
  \bibinfo {pages} {25001} (\bibinfo {year} {2020})}\BibitemShut {NoStop}%
\bibitem [{\citenamefont {Pashinin}\ and\ \citenamefont
  {Fedorov}(1978)}]{Pashinin1978}%
  \BibitemOpen
  \bibfield  {author} {\bibinfo {author} {\bibfnamefont {P.}~\bibnamefont
  {Pashinin}}\ and\ \bibinfo {author} {\bibfnamefont {M.}~\bibnamefont
  {Fedorov}},\ }\href
  {http://www.jetp.ras.ru/cgi-bin/e/index/e/48/2/p228?a=list} {\bibfield
  {journal} {\bibinfo  {journal} {J. Exp. Theor. Phys.}\ }\textbf {\bibinfo
  {volume} {48}},\ \bibinfo {pages} {228} (\bibinfo {year} {1978})}\BibitemShut
  {NoStop}%
\bibitem [{\citenamefont {Munirov}\ and\ \citenamefont
  {Fisch}(2017{\natexlab{a}})}]{Munirov2017b}%
  \BibitemOpen
  \bibfield  {author} {\bibinfo {author} {\bibfnamefont {V.~R.}\ \bibnamefont
  {Munirov}}\ and\ \bibinfo {author} {\bibfnamefont {N.~J.}\ \bibnamefont
  {Fisch}},\ }\href {https://doi.org/10.1103/PhysRevE.96.053211} {\bibfield
  {journal} {\bibinfo  {journal} {Phys. Rev. E}\ }\textbf {\bibinfo {volume}
  {96}},\ \bibinfo {pages} {053211} (\bibinfo {year}
  {2017}{\natexlab{a}})}\BibitemShut {NoStop}%
\bibitem [{\citenamefont {Bunkin}\ and\ \citenamefont
  {Fedorov}(1966)}]{Bunkin1966}%
  \BibitemOpen
  \bibfield  {author} {\bibinfo {author} {\bibfnamefont {F.}~\bibnamefont
  {Bunkin}}\ and\ \bibinfo {author} {\bibfnamefont {M.}~\bibnamefont
  {Fedorov}},\ }\href
  {http://www.jetp.ras.ru/cgi-bin/e/index/e/22/4/p844?a=list} {\bibfield
  {journal} {\bibinfo  {journal} {J. Exp. Theor. Phys.}\ }\textbf {\bibinfo
  {volume} {22}},\ \bibinfo {pages} {844} (\bibinfo {year} {1966})}\BibitemShut
  {NoStop}%
\bibitem [{\citenamefont {Denisov}\ and\ \citenamefont
  {Fedorov}(1968)}]{Denisov1968}%
  \BibitemOpen
  \bibfield  {author} {\bibinfo {author} {\bibfnamefont {M.}~\bibnamefont
  {Denisov}}\ and\ \bibinfo {author} {\bibfnamefont {M.}~\bibnamefont
  {Fedorov}},\ }\href
  {http://www.jetp.ras.ru/cgi-bin/e/index/e/26/4/p779?a=list} {\bibfield
  {journal} {\bibinfo  {journal} {J. Exp. Theor. Phys.}\ }\textbf {\bibinfo
  {volume} {26}},\ \bibinfo {pages} {779} (\bibinfo {year} {1968})}\BibitemShut
  {NoStop}%
\bibitem [{\citenamefont {Schlessinger}\ and\ \citenamefont
  {Wright}(1979)}]{Schlessinger1979}%
  \BibitemOpen
  \bibfield  {author} {\bibinfo {author} {\bibfnamefont {L.}~\bibnamefont
  {Schlessinger}}\ and\ \bibinfo {author} {\bibfnamefont {J.}~\bibnamefont
  {Wright}},\ }\href {https://doi.org/10.1103/PhysRevA.20.1934} {\bibfield
  {journal} {\bibinfo  {journal} {Phys. Rev. A}\ }\textbf {\bibinfo {volume}
  {20}},\ \bibinfo {pages} {1934} (\bibinfo {year} {1979})}\BibitemShut
  {NoStop}%
\bibitem [{\citenamefont {Pert}(1995)}]{Pert1995}%
  \BibitemOpen
  \bibfield  {author} {\bibinfo {author} {\bibfnamefont {G.~J.}\ \bibnamefont
  {Pert}},\ }\href {https://doi.org/10.1103/PhysRevE.51.4778} {\bibfield
  {journal} {\bibinfo  {journal} {Phys. Rev. E}\ }\textbf {\bibinfo {volume}
  {51}},\ \bibinfo {pages} {4778} (\bibinfo {year} {1995})}\BibitemShut
  {NoStop}%
\bibitem [{\citenamefont {Shvets}\ and\ \citenamefont
  {Fisch}(1997)}]{Shvets1997}%
  \BibitemOpen
  \bibfield  {author} {\bibinfo {author} {\bibfnamefont {G.}~\bibnamefont
  {Shvets}}\ and\ \bibinfo {author} {\bibfnamefont {N.~J.}\ \bibnamefont
  {Fisch}},\ }\href {https://doi.org/10.1063/1.872101} {\bibfield  {journal}
  {\bibinfo  {journal} {Phys. Plasmas}\ }\textbf {\bibinfo {volume} {4}},\
  \bibinfo {pages} {428} (\bibinfo {year} {1997})}\BibitemShut {NoStop}%
\bibitem [{\citenamefont {Balakin}\ and\ \citenamefont
  {Fraiman}(2001)}]{Balakin2001}%
  \BibitemOpen
  \bibfield  {author} {\bibinfo {author} {\bibfnamefont {A.~A.}\ \bibnamefont
  {Balakin}}\ and\ \bibinfo {author} {\bibfnamefont {G.~M.}\ \bibnamefont
  {Fraiman}},\ }\href {https://doi.org/10.1134/1.1420438} {\bibfield  {journal}
  {\bibinfo  {journal} {J. Exp. Theor. Phys.}\ }\textbf {\bibinfo {volume}
  {93}},\ \bibinfo {pages} {695} (\bibinfo {year} {2001})}\BibitemShut
  {NoStop}%
\bibitem [{\citenamefont {Brantov}\ \emph {et~al.}(2003)\citenamefont
  {Brantov}, \citenamefont {Rozmus}, \citenamefont {Sydora}, \citenamefont
  {Capjack}, \citenamefont {Bychenkov},\ and\ \citenamefont
  {Tikhonchuk}}]{Brantov2003}%
  \BibitemOpen
  \bibfield  {author} {\bibinfo {author} {\bibfnamefont {A.}~\bibnamefont
  {Brantov}}, \bibinfo {author} {\bibfnamefont {W.}~\bibnamefont {Rozmus}},
  \bibinfo {author} {\bibfnamefont {R.}~\bibnamefont {Sydora}}, \bibinfo
  {author} {\bibfnamefont {C.~E.}\ \bibnamefont {Capjack}}, \bibinfo {author}
  {\bibfnamefont {V.~Y.}\ \bibnamefont {Bychenkov}},\ and\ \bibinfo {author}
  {\bibfnamefont {V.~T.}\ \bibnamefont {Tikhonchuk}},\ }\href
  {https://doi.org/10.1063/1.1586917} {\bibfield  {journal} {\bibinfo
  {journal} {Phys. Plasmas}\ }\textbf {\bibinfo {volume} {10}},\ \bibinfo
  {pages} {3385} (\bibinfo {year} {2003})}\BibitemShut {NoStop}%
\bibitem [{\citenamefont {{Krolik}}\ and\ \citenamefont
  {{Kallman}}(1984)}]{Krolik1984}%
  \BibitemOpen
  \bibfield  {author} {\bibinfo {author} {\bibfnamefont {J.~H.}\ \bibnamefont
  {{Krolik}}}\ and\ \bibinfo {author} {\bibfnamefont {T.~R.}\ \bibnamefont
  {{Kallman}}},\ }\href {https://doi.org/10.1086/162608} {\bibfield  {journal}
  {\bibinfo  {journal} {Astrophys. J.}\ }\textbf {\bibinfo {volume} {286}},\
  \bibinfo {pages} {366} (\bibinfo {year} {1984})}\BibitemShut {NoStop}%
\bibitem [{\citenamefont {Mihajlov}\ \emph {et~al.}(2015)\citenamefont
  {Mihajlov}, \citenamefont {Sre{\'c}kovi{\'c}},\ and\ \citenamefont
  {Sakan}}]{Mihajlov2015}%
  \BibitemOpen
  \bibfield  {author} {\bibinfo {author} {\bibfnamefont {A.~A.}\ \bibnamefont
  {Mihajlov}}, \bibinfo {author} {\bibfnamefont {V.~A.}\ \bibnamefont
  {Sre{\'c}kovi{\'c}}},\ and\ \bibinfo {author} {\bibfnamefont {N.~M.}\
  \bibnamefont {Sakan}},\ }\href {https://doi.org/10.1007/s12036-015-9350-0}
  {\bibfield  {journal} {\bibinfo  {journal} {J. Astrophys. Astron.}\ }\textbf
  {\bibinfo {volume} {36}},\ \bibinfo {pages} {0} (\bibinfo {year}
  {2015})}\BibitemShut {NoStop}%
\bibitem [{\citenamefont {{Itoh}}\ \emph {et~al.}(1985)\citenamefont {{Itoh}},
  \citenamefont {{Nakagawa}},\ and\ \citenamefont {{Kohyama}}}]{Itoh1985}%
  \BibitemOpen
  \bibfield  {author} {\bibinfo {author} {\bibfnamefont {N.}~\bibnamefont
  {{Itoh}}}, \bibinfo {author} {\bibfnamefont {M.}~\bibnamefont {{Nakagawa}}},\
  and\ \bibinfo {author} {\bibfnamefont {Y.}~\bibnamefont {{Kohyama}}},\ }\href
  {https://doi.org/10.1086/163269} {\bibfield  {journal} {\bibinfo  {journal}
  {Astrophys. J.}\ }\textbf {\bibinfo {volume} {294}},\ \bibinfo {pages} {17}
  (\bibinfo {year} {1985})}\BibitemShut {NoStop}%
\bibitem [{\citenamefont {Dimitrijevi{\'c}}\ \emph {et~al.}(2018)\citenamefont
  {Dimitrijevi{\'c}}, \citenamefont {Sre{\'c}kovi{\'c}}, \citenamefont {Sakan},
  \citenamefont {Bezuglov},\ and\ \citenamefont
  {Klyucharev}}]{Dimitrijevic2018}%
  \BibitemOpen
  \bibfield  {author} {\bibinfo {author} {\bibfnamefont {M.~S.}\ \bibnamefont
  {Dimitrijevi{\'c}}}, \bibinfo {author} {\bibfnamefont {V.~A.}\ \bibnamefont
  {Sre{\'c}kovi{\'c}}}, \bibinfo {author} {\bibfnamefont {N.~M.}\ \bibnamefont
  {Sakan}}, \bibinfo {author} {\bibfnamefont {N.~N.}\ \bibnamefont
  {Bezuglov}},\ and\ \bibinfo {author} {\bibfnamefont {A.~N.}\ \bibnamefont
  {Klyucharev}},\ }\href {https://doi.org/10.1134/S0016793218080054} {\bibfield
   {journal} {\bibinfo  {journal} {Geomagn. Aeron.}\ }\textbf {\bibinfo
  {volume} {58}},\ \bibinfo {pages} {1067} (\bibinfo {year}
  {2018})}\BibitemShut {NoStop}%
\bibitem [{\citenamefont {Nozawa}\ \emph {et~al.}(1998)\citenamefont {Nozawa},
  \citenamefont {Itoh},\ and\ \citenamefont {Kohyama}}]{Nozawa1998}%
  \BibitemOpen
  \bibfield  {author} {\bibinfo {author} {\bibfnamefont {S.}~\bibnamefont
  {Nozawa}}, \bibinfo {author} {\bibfnamefont {N.}~\bibnamefont {Itoh}},\ and\
  \bibinfo {author} {\bibfnamefont {Y.}~\bibnamefont {Kohyama}},\ }\href
  {https://doi.org/10.1086/306352} {\bibfield  {journal} {\bibinfo  {journal}
  {Astrophys. J.}\ }\textbf {\bibinfo {volume} {507}},\ \bibinfo {pages} {530}
  (\bibinfo {year} {1998})}\BibitemShut {NoStop}%
\bibitem [{\citenamefont {Itoh}\ \emph {et~al.}(2000)\citenamefont {Itoh},
  \citenamefont {Sakamoto}, \citenamefont {Kusano}, \citenamefont {Nozawa},\
  and\ \citenamefont {Kohyama}}]{Itoh2000}%
  \BibitemOpen
  \bibfield  {author} {\bibinfo {author} {\bibfnamefont {N.}~\bibnamefont
  {Itoh}}, \bibinfo {author} {\bibfnamefont {T.}~\bibnamefont {Sakamoto}},
  \bibinfo {author} {\bibfnamefont {S.}~\bibnamefont {Kusano}}, \bibinfo
  {author} {\bibfnamefont {S.}~\bibnamefont {Nozawa}},\ and\ \bibinfo {author}
  {\bibfnamefont {Y.}~\bibnamefont {Kohyama}},\ }\href
  {https://doi.org/10.1086/313375} {\bibfield  {journal} {\bibinfo  {journal}
  {Astrophys. J., Suppl. Ser.}\ }\textbf {\bibinfo {volume} {128}},\ \bibinfo
  {pages} {125} (\bibinfo {year} {2000})}\BibitemShut {NoStop}%
\bibitem [{\citenamefont {Itoh}\ \emph {et~al.}(2002)\citenamefont {Itoh},
  \citenamefont {Sakamoto}, \citenamefont {Kusano}, \citenamefont {Kawana},\
  and\ \citenamefont {Nozawa}}]{Itoh2002}%
  \BibitemOpen
  \bibfield  {author} {\bibinfo {author} {\bibfnamefont {N.}~\bibnamefont
  {Itoh}}, \bibinfo {author} {\bibfnamefont {T.}~\bibnamefont {Sakamoto}},
  \bibinfo {author} {\bibfnamefont {S.}~\bibnamefont {Kusano}}, \bibinfo
  {author} {\bibfnamefont {Y.}~\bibnamefont {Kawana}},\ and\ \bibinfo {author}
  {\bibfnamefont {S.}~\bibnamefont {Nozawa}},\ }\href
  {https://doi.org/10.1051/0004-6361:20011664} {\bibfield  {journal} {\bibinfo
  {journal} {Astron. Astrophys.}\ }\textbf {\bibinfo {volume} {382}},\ \bibinfo
  {pages} {722} (\bibinfo {year} {2002})}\BibitemShut {NoStop}%
\bibitem [{\citenamefont {Uchiyama}\ \emph {et~al.}(2002)\citenamefont
  {Uchiyama}, \citenamefont {Takahashi}, \citenamefont {Aharonian},\ and\
  \citenamefont {Mattox}}]{Uchiyama2002}%
  \BibitemOpen
  \bibfield  {author} {\bibinfo {author} {\bibfnamefont {Y.}~\bibnamefont
  {Uchiyama}}, \bibinfo {author} {\bibfnamefont {T.}~\bibnamefont {Takahashi}},
  \bibinfo {author} {\bibfnamefont {F.~A.}\ \bibnamefont {Aharonian}},\ and\
  \bibinfo {author} {\bibfnamefont {J.~R.}\ \bibnamefont {Mattox}},\ }\href
  {https://doi.org/10.1086/340121} {\bibfield  {journal} {\bibinfo  {journal}
  {Astrophys. J.}\ }\textbf {\bibinfo {volume} {571}},\ \bibinfo {pages} {866}
  (\bibinfo {year} {2002})}\BibitemShut {NoStop}%
\bibitem [{\citenamefont {Vink}(2008)}]{Vink2008}%
  \BibitemOpen
  \bibfield  {author} {\bibinfo {author} {\bibfnamefont {J.}~\bibnamefont
  {Vink}},\ }\href {https://doi.org/10.1051/0004-6361:200809669} {\bibfield
  {journal} {\bibinfo  {journal} {Astron. Astrophys.}\ }\textbf {\bibinfo
  {volume} {486}},\ \bibinfo {pages} {837} (\bibinfo {year}
  {2008})}\BibitemShut {NoStop}%
\bibitem [{\citenamefont {Fang}\ and\ \citenamefont {Zhang}(2008)}]{Fang2008}%
  \BibitemOpen
  \bibfield  {author} {\bibinfo {author} {\bibfnamefont {J.}~\bibnamefont
  {Fang}}\ and\ \bibinfo {author} {\bibfnamefont {L.}~\bibnamefont {Zhang}},\
  }\href {https://doi.org/10.1111/j.1365-2966.2007.12766.x} {\bibfield
  {journal} {\bibinfo  {journal} {Mon. Not. R. Astron. Soc.}\ }\textbf
  {\bibinfo {volume} {384}},\ \bibinfo {pages} {1119} (\bibinfo {year}
  {2008})}\BibitemShut {NoStop}%
\bibitem [{\citenamefont {Langdon}(1980)}]{Langdon1980}%
  \BibitemOpen
  \bibfield  {author} {\bibinfo {author} {\bibfnamefont {A.~B.}\ \bibnamefont
  {Langdon}},\ }\href {https://doi.org/10.1103/PhysRevLett.44.575} {\bibfield
  {journal} {\bibinfo  {journal} {Phys. Rev. Lett.}\ }\textbf {\bibinfo
  {volume} {44}},\ \bibinfo {pages} {575} (\bibinfo {year} {1980})}\BibitemShut
  {NoStop}%
\bibitem [{\citenamefont {Matte}\ \emph {et~al.}(1988)\citenamefont {Matte},
  \citenamefont {Lamoureux}, \citenamefont {Moller}, \citenamefont {Yin},
  \citenamefont {Delettrez}, \citenamefont {Virmont},\ and\ \citenamefont
  {Johnston}}]{Matte1988}%
  \BibitemOpen
  \bibfield  {author} {\bibinfo {author} {\bibfnamefont {J.~P.}\ \bibnamefont
  {Matte}}, \bibinfo {author} {\bibfnamefont {M.}~\bibnamefont {Lamoureux}},
  \bibinfo {author} {\bibfnamefont {C.}~\bibnamefont {Moller}}, \bibinfo
  {author} {\bibfnamefont {R.~Y.}\ \bibnamefont {Yin}}, \bibinfo {author}
  {\bibfnamefont {J.}~\bibnamefont {Delettrez}}, \bibinfo {author}
  {\bibfnamefont {J.}~\bibnamefont {Virmont}},\ and\ \bibinfo {author}
  {\bibfnamefont {T.~W.}\ \bibnamefont {Johnston}},\ }\href
  {https://doi.org/10.1088/0741-3335/30/12/004} {\bibfield  {journal} {\bibinfo
   {journal} {Plasma Phys. Control. Fusion}\ }\textbf {\bibinfo {volume}
  {30}},\ \bibinfo {pages} {1665} (\bibinfo {year} {1988})}\BibitemShut
  {NoStop}%
\bibitem [{\citenamefont {Dubroca}\ \emph {et~al.}(2004)\citenamefont
  {Dubroca}, \citenamefont {Tchong}, \citenamefont {Charrier}, \citenamefont
  {Tikhonchuk},\ and\ \citenamefont {Morreeuw}}]{Dubroca2004}%
  \BibitemOpen
  \bibfield  {author} {\bibinfo {author} {\bibfnamefont {B.}~\bibnamefont
  {Dubroca}}, \bibinfo {author} {\bibfnamefont {M.}~\bibnamefont {Tchong}},
  \bibinfo {author} {\bibfnamefont {P.}~\bibnamefont {Charrier}}, \bibinfo
  {author} {\bibfnamefont {V.~T.}\ \bibnamefont {Tikhonchuk}},\ and\ \bibinfo
  {author} {\bibfnamefont {J.-P.}\ \bibnamefont {Morreeuw}},\ }\href
  {https://doi.org/10.1063/1.1760089} {\bibfield  {journal} {\bibinfo
  {journal} {Phys. Plasmas}\ }\textbf {\bibinfo {volume} {11}},\ \bibinfo
  {pages} {3830} (\bibinfo {year} {2004})}\BibitemShut {NoStop}%
\bibitem [{\citenamefont {Munirov}\ and\ \citenamefont
  {Fisch}(2017{\natexlab{b}})}]{Munirov2017a}%
  \BibitemOpen
  \bibfield  {author} {\bibinfo {author} {\bibfnamefont {V.~R.}\ \bibnamefont
  {Munirov}}\ and\ \bibinfo {author} {\bibfnamefont {N.~J.}\ \bibnamefont
  {Fisch}},\ }\href {https://doi.org/10.1103/PhysRevE.95.013205} {\bibfield
  {journal} {\bibinfo  {journal} {Phys. Rev. E}\ }\textbf {\bibinfo {volume}
  {95}},\ \bibinfo {pages} {013205} (\bibinfo {year}
  {2017}{\natexlab{b}})}\BibitemShut {NoStop}%
\bibitem [{\citenamefont {Munirov}\ and\ \citenamefont
  {Fisch}(2019)}]{Munirov2019}%
  \BibitemOpen
  \bibfield  {author} {\bibinfo {author} {\bibfnamefont {V.~R.}\ \bibnamefont
  {Munirov}}\ and\ \bibinfo {author} {\bibfnamefont {N.~J.}\ \bibnamefont
  {Fisch}},\ }\href {https://doi.org/10.1103/PhysRevE.100.023202} {\bibfield
  {journal} {\bibinfo  {journal} {Phys. Rev. E}\ }\textbf {\bibinfo {volume}
  {100}},\ \bibinfo {pages} {023202} (\bibinfo {year} {2019})}\BibitemShut
  {NoStop}%
\bibitem [{\citenamefont {Ochs}\ and\ \citenamefont {Fisch}(2020)}]{Ochs2020}%
  \BibitemOpen
  \bibfield  {author} {\bibinfo {author} {\bibfnamefont {I.~E.}\ \bibnamefont
  {Ochs}}\ and\ \bibinfo {author} {\bibfnamefont {N.~J.}\ \bibnamefont
  {Fisch}},\ }\href {https://doi.org/10.3847/1538-4357/abc4e8} {\bibfield
  {journal} {\bibinfo  {journal} {Astrophys. J.}\ }\textbf {\bibinfo {volume}
  {905}},\ \bibinfo {pages} {13} (\bibinfo {year} {2020})}\BibitemShut
  {NoStop}%
\bibitem [{\citenamefont {Shohet}(1968)}]{Shohet1968}%
  \BibitemOpen
  \bibfield  {author} {\bibinfo {author} {\bibfnamefont {J.~L.}\ \bibnamefont
  {Shohet}},\ }\href {https://doi.org/10.1063/1.1692044} {\bibfield  {journal}
  {\bibinfo  {journal} {Phys. Fluids}\ }\textbf {\bibinfo {volume} {11}},\
  \bibinfo {pages} {1065} (\bibinfo {year} {1968})}\BibitemShut {NoStop}%
\bibitem [{\citenamefont {{Dermer}}\ and\ \citenamefont
  {{Ramaty}}(1986)}]{Dermer1986}%
  \BibitemOpen
  \bibfield  {author} {\bibinfo {author} {\bibfnamefont {C.~D.}\ \bibnamefont
  {{Dermer}}}\ and\ \bibinfo {author} {\bibfnamefont {R.}~\bibnamefont
  {{Ramaty}}},\ }\href {https://doi.org/10.1086/163959} {\bibfield  {journal}
  {\bibinfo  {journal} {Astrophys. J.}\ }\textbf {\bibinfo {volume} {301}},\
  \bibinfo {pages} {962} (\bibinfo {year} {1986})}\BibitemShut {NoStop}%
\bibitem [{\citenamefont {Ferrante}\ \emph {et~al.}(2001)\citenamefont
  {Ferrante}, \citenamefont {Zarcone},\ and\ \citenamefont
  {Uryupin}}]{Ferrante2001}%
  \BibitemOpen
  \bibfield  {author} {\bibinfo {author} {\bibfnamefont {G.}~\bibnamefont
  {Ferrante}}, \bibinfo {author} {\bibfnamefont {M.}~\bibnamefont {Zarcone}},\
  and\ \bibinfo {author} {\bibfnamefont {S.~A.}\ \bibnamefont {Uryupin}},\
  }\href {https://doi.org/10.1063/1.1405014} {\bibfield  {journal} {\bibinfo
  {journal} {Phys. Plasmas}\ }\textbf {\bibinfo {volume} {8}},\ \bibinfo
  {pages} {4745} (\bibinfo {year} {2001})}\BibitemShut {NoStop}%
\bibitem [{\citenamefont {Massone}\ \emph {et~al.}(2004)\citenamefont
  {Massone}, \citenamefont {Emslie}, \citenamefont {Kontar}, \citenamefont
  {Piana}, \citenamefont {Prato},\ and\ \citenamefont {Brown}}]{Massone2004}%
  \BibitemOpen
  \bibfield  {author} {\bibinfo {author} {\bibfnamefont {A.~M.}\ \bibnamefont
  {Massone}}, \bibinfo {author} {\bibfnamefont {A.~G.}\ \bibnamefont {Emslie}},
  \bibinfo {author} {\bibfnamefont {E.~P.}\ \bibnamefont {Kontar}}, \bibinfo
  {author} {\bibfnamefont {M.}~\bibnamefont {Piana}}, \bibinfo {author}
  {\bibfnamefont {M.}~\bibnamefont {Prato}},\ and\ \bibinfo {author}
  {\bibfnamefont {J.~C.}\ \bibnamefont {Brown}},\ }\href
  {https://doi.org/10.1086/423127} {\bibfield  {journal} {\bibinfo  {journal}
  {Astrophys. J.}\ }\textbf {\bibinfo {volume} {613}},\ \bibinfo {pages} {1233}
  (\bibinfo {year} {2004})}\BibitemShut {NoStop}%
\bibitem [{\citenamefont {Oparin}\ \emph {et~al.}(2020)\citenamefont {Oparin},
  \citenamefont {Charikov}, \citenamefont {Ovchinnikova},\ and\ \citenamefont
  {Shabalin}}]{Oparin2020}%
  \BibitemOpen
  \bibfield  {author} {\bibinfo {author} {\bibfnamefont {I.~D.}\ \bibnamefont
  {Oparin}}, \bibinfo {author} {\bibfnamefont {Y.~E.}\ \bibnamefont
  {Charikov}}, \bibinfo {author} {\bibfnamefont {E.~P.}\ \bibnamefont
  {Ovchinnikova}},\ and\ \bibinfo {author} {\bibfnamefont {A.~N.}\ \bibnamefont
  {Shabalin}},\ }\href {https://doi.org/10.1134/S0016793220070191} {\bibfield
  {journal} {\bibinfo  {journal} {Geomagn. Aeron.}\ }\textbf {\bibinfo {volume}
  {60}},\ \bibinfo {pages} {889} (\bibinfo {year} {2020})}\BibitemShut
  {NoStop}%
\bibitem [{\citenamefont {{Dawson}}(1983)}]{Dawson1983}%
  \BibitemOpen
  \bibfield  {author} {\bibinfo {author} {\bibfnamefont {J.~M.}\ \bibnamefont
  {{Dawson}}},\ }\href {https://ui.adsabs.harvard.edu/abs/1983naja.rept...20D}
  {\bibinfo {title} {{Advanced Fusion Reactors}}},\ \bibinfo {howpublished}
  {Lectures held at Nagoya, Japan, 20-22 Oct. 1982} (\bibinfo {year}
  {1983})\BibitemShut {NoStop}%
\bibitem [{\citenamefont {Davidson}\ \emph {et~al.}(1979)\citenamefont
  {Davidson}, \citenamefont {Berg}, \citenamefont {Lowry}, \citenamefont
  {Dwarakanath}, \citenamefont {Sierk},\ and\ \citenamefont
  {Batay-Csorba}}]{Davidson1979}%
  \BibitemOpen
  \bibfield  {author} {\bibinfo {author} {\bibfnamefont {J.}~\bibnamefont
  {Davidson}}, \bibinfo {author} {\bibfnamefont {H.}~\bibnamefont {Berg}},
  \bibinfo {author} {\bibfnamefont {M.}~\bibnamefont {Lowry}}, \bibinfo
  {author} {\bibfnamefont {M.}~\bibnamefont {Dwarakanath}}, \bibinfo {author}
  {\bibfnamefont {A.}~\bibnamefont {Sierk}},\ and\ \bibinfo {author}
  {\bibfnamefont {P.}~\bibnamefont {Batay-Csorba}},\ }\href
  {https://doi.org/https://doi.org/10.1016/0375-9474(79)90647-X} {\bibfield
  {journal} {\bibinfo  {journal} {Nucl. Phys. A}\ }\textbf {\bibinfo {volume}
  {315}},\ \bibinfo {pages} {253} (\bibinfo {year} {1979})}\BibitemShut
  {NoStop}%
\bibitem [{\citenamefont {Wurzel}\ and\ \citenamefont
  {Hsu}(2022)}]{Wurzel2022}%
  \BibitemOpen
  \bibfield  {author} {\bibinfo {author} {\bibfnamefont {S.~E.}\ \bibnamefont
  {Wurzel}}\ and\ \bibinfo {author} {\bibfnamefont {S.~C.}\ \bibnamefont
  {Hsu}},\ }\href {https://doi.org/10.1063/5.0083990} {\bibfield  {journal}
  {\bibinfo  {journal} {Phys. Plasmas}\ }\textbf {\bibinfo {volume} {29}},\
  \bibinfo {pages} {062103} (\bibinfo {year} {2022})}\BibitemShut {NoStop}%
\bibitem [{\citenamefont {Rider}(1995)}]{Rider1995}%
  \BibitemOpen
  \bibfield  {author} {\bibinfo {author} {\bibfnamefont {T.~H.}\ \bibnamefont
  {Rider}},\ }\href {https://doi.org/10.1063/1.871273} {\bibfield  {journal}
  {\bibinfo  {journal} {Phys. Plasmas}\ }\textbf {\bibinfo {volume} {2}},\
  \bibinfo {pages} {1853} (\bibinfo {year} {1995})}\BibitemShut {NoStop}%
\bibitem [{\citenamefont {Rider}(1997)}]{Rider1997}%
  \BibitemOpen
  \bibfield  {author} {\bibinfo {author} {\bibfnamefont {T.~H.}\ \bibnamefont
  {Rider}},\ }\href {https://doi.org/10.1063/1.872556} {\bibfield  {journal}
  {\bibinfo  {journal} {Phys. Plasmas}\ }\textbf {\bibinfo {volume} {4}},\
  \bibinfo {pages} {1039} (\bibinfo {year} {1997})}\BibitemShut {NoStop}%
\bibitem [{\citenamefont {Nevins}(1998)}]{Nevins1998}%
  \BibitemOpen
  \bibfield  {author} {\bibinfo {author} {\bibfnamefont {W.~M.}\ \bibnamefont
  {Nevins}},\ }\href {https://doi.org/10.1023/A:1022513215080} {\bibfield
  {journal} {\bibinfo  {journal} {J. Fusion Energy}\ }\textbf {\bibinfo
  {volume} {17}},\ \bibinfo {pages} {25} (\bibinfo {year} {1998})}\BibitemShut
  {NoStop}%
\bibitem [{\citenamefont {Sikora}\ and\ \citenamefont
  {Weller}(2016)}]{Sikora2016}%
  \BibitemOpen
  \bibfield  {author} {\bibinfo {author} {\bibfnamefont {M.~H.}\ \bibnamefont
  {Sikora}}\ and\ \bibinfo {author} {\bibfnamefont {H.~R.}\ \bibnamefont
  {Weller}},\ }\href {https://doi.org/10.1007/s10894-016-0069-y} {\bibfield
  {journal} {\bibinfo  {journal} {J. Fusion Energy}\ }\textbf {\bibinfo
  {volume} {35}},\ \bibinfo {pages} {538} (\bibinfo {year} {2016})}\BibitemShut
  {NoStop}%
\bibitem [{\citenamefont {Fisch}\ and\ \citenamefont
  {Rax}(1992{\natexlab{a}})}]{Fisch1992a}%
  \BibitemOpen
  \bibfield  {author} {\bibinfo {author} {\bibfnamefont {N.~J.}\ \bibnamefont
  {Fisch}}\ and\ \bibinfo {author} {\bibfnamefont {J.-M.}\ \bibnamefont
  {Rax}},\ }\href {https://doi.org/10.1103/PhysRevLett.69.612} {\bibfield
  {journal} {\bibinfo  {journal} {Phys. Rev. Lett.}\ }\textbf {\bibinfo
  {volume} {69}},\ \bibinfo {pages} {612} (\bibinfo {year}
  {1992}{\natexlab{a}})}\BibitemShut {NoStop}%
\bibitem [{\citenamefont {Fisch}\ and\ \citenamefont
  {Rax}(1992{\natexlab{b}})}]{Fisch1992b}%
  \BibitemOpen
  \bibfield  {author} {\bibinfo {author} {\bibfnamefont {N.}~\bibnamefont
  {Fisch}}\ and\ \bibinfo {author} {\bibfnamefont {J.-M.}\ \bibnamefont
  {Rax}},\ }\href {https://doi.org/10.1088/0029-5515/32/4/I02} {\bibfield
  {journal} {\bibinfo  {journal} {Nucl. Fusion}\ }\textbf {\bibinfo {volume}
  {32}},\ \bibinfo {pages} {549} (\bibinfo {year}
  {1992}{\natexlab{b}})}\BibitemShut {NoStop}%
\bibitem [{\citenamefont {Hay}\ and\ \citenamefont {Fisch}(2015)}]{Hay2015}%
  \BibitemOpen
  \bibfield  {author} {\bibinfo {author} {\bibfnamefont {M.~J.}\ \bibnamefont
  {Hay}}\ and\ \bibinfo {author} {\bibfnamefont {N.~J.}\ \bibnamefont
  {Fisch}},\ }\href {https://doi.org/10.1063/1.4936346} {\bibfield  {journal}
  {\bibinfo  {journal} {Phys. Plasmas}\ }\textbf {\bibinfo {volume} {22}},\
  \bibinfo {pages} {112116} (\bibinfo {year} {2015})}\BibitemShut {NoStop}%
\bibitem [{\citenamefont {Putvinski}\ \emph {et~al.}(2019)\citenamefont
  {Putvinski}, \citenamefont {Ryutov},\ and\ \citenamefont
  {Yushmanov}}]{Putvinski2019}%
  \BibitemOpen
  \bibfield  {author} {\bibinfo {author} {\bibfnamefont {S.}~\bibnamefont
  {Putvinski}}, \bibinfo {author} {\bibfnamefont {D.}~\bibnamefont {Ryutov}},\
  and\ \bibinfo {author} {\bibfnamefont {P.}~\bibnamefont {Yushmanov}},\ }\href
  {https://doi.org/10.1088/1741-4326/ab1a60} {\bibfield  {journal} {\bibinfo
  {journal} {Nucl. Fusion}\ }\textbf {\bibinfo {volume} {59}},\ \bibinfo
  {pages} {076018} (\bibinfo {year} {2019})}\BibitemShut {NoStop}%
\bibitem [{\citenamefont {Ochs}\ \emph {et~al.}(2022)\citenamefont {Ochs},
  \citenamefont {Kolmes}, \citenamefont {Mlodik}, \citenamefont {Rubin},\ and\
  \citenamefont {Fisch}}]{Ochs_pB11_2022}%
  \BibitemOpen
  \bibfield  {author} {\bibinfo {author} {\bibfnamefont {I.~E.}\ \bibnamefont
  {Ochs}}, \bibinfo {author} {\bibfnamefont {E.~J.}\ \bibnamefont {Kolmes}},
  \bibinfo {author} {\bibfnamefont {M.~E.}\ \bibnamefont {Mlodik}}, \bibinfo
  {author} {\bibfnamefont {T.}~\bibnamefont {Rubin}},\ and\ \bibinfo {author}
  {\bibfnamefont {N.~J.}\ \bibnamefont {Fisch}},\ }\href
  {https://doi.org/10.1103/PhysRevE.106.055215} {\bibfield  {journal} {\bibinfo
   {journal} {Phys. Rev. E}\ }\textbf {\bibinfo {volume} {106}},\ \bibinfo
  {pages} {055215} (\bibinfo {year} {2022})}\BibitemShut {NoStop}%
\bibitem [{\citenamefont {Kolmes}\ \emph {et~al.}(2022)\citenamefont {Kolmes},
  \citenamefont {Ochs},\ and\ \citenamefont {Fisch}}]{Kolmes2022}%
  \BibitemOpen
  \bibfield  {author} {\bibinfo {author} {\bibfnamefont {E.~J.}\ \bibnamefont
  {Kolmes}}, \bibinfo {author} {\bibfnamefont {I.~E.}\ \bibnamefont {Ochs}},\
  and\ \bibinfo {author} {\bibfnamefont {N.~J.}\ \bibnamefont {Fisch}},\ }\href
  {https://doi.org/10.1063/5.0119434} {\bibfield  {journal} {\bibinfo
  {journal} {Phys. Plasmas}\ }\textbf {\bibinfo {volume} {29}},\ \bibinfo
  {pages} {110701} (\bibinfo {year} {2022})}\BibitemShut {NoStop}%
\bibitem [{\citenamefont {Kurilenkov}\ \emph {et~al.}(2021)\citenamefont
  {Kurilenkov}, \citenamefont {Oginov}, \citenamefont {Tarakanov},
  \citenamefont {Gus'kov},\ and\ \citenamefont {Samoylov}}]{Kurilenkov2021}%
  \BibitemOpen
  \bibfield  {author} {\bibinfo {author} {\bibfnamefont {Y.~K.}\ \bibnamefont
  {Kurilenkov}}, \bibinfo {author} {\bibfnamefont {A.~V.}\ \bibnamefont
  {Oginov}}, \bibinfo {author} {\bibfnamefont {V.~P.}\ \bibnamefont
  {Tarakanov}}, \bibinfo {author} {\bibfnamefont {S.~Y.}\ \bibnamefont
  {Gus'kov}},\ and\ \bibinfo {author} {\bibfnamefont {I.~S.}\ \bibnamefont
  {Samoylov}},\ }\href {https://doi.org/10.1103/PhysRevE.103.043208} {\bibfield
   {journal} {\bibinfo  {journal} {Phys. Rev. E}\ }\textbf {\bibinfo {volume}
  {103}},\ \bibinfo {pages} {043208} (\bibinfo {year} {2021})}\BibitemShut
  {NoStop}%
\bibitem [{\citenamefont {Lerner}\ \emph {et~al.}(2023)\citenamefont {Lerner},
  \citenamefont {Hassan}, \citenamefont {Karamitsos-Zivkovic},\ and\
  \citenamefont {Fritsch}}]{Lerner2023}%
  \BibitemOpen
  \bibfield  {author} {\bibinfo {author} {\bibfnamefont {E.~J.}\ \bibnamefont
  {Lerner}}, \bibinfo {author} {\bibfnamefont {S.~M.}\ \bibnamefont {Hassan}},
  \bibinfo {author} {\bibfnamefont {I.}~\bibnamefont {Karamitsos-Zivkovic}},\
  and\ \bibinfo {author} {\bibfnamefont {R.}~\bibnamefont {Fritsch}},\ }\href
  {https://doi.org/10.1007/s10894-023-00345-z} {\bibfield  {journal} {\bibinfo
  {journal} {J. Fusion Energy}\ }\textbf {\bibinfo {volume} {42}},\ \bibinfo
  {pages} {7} (\bibinfo {year} {2023})}\BibitemShut {NoStop}%
\bibitem [{\citenamefont {Magee}\ \emph {et~al.}(2023)\citenamefont {Magee},
  \citenamefont {Ogawa}, \citenamefont {Tajima}, \citenamefont {Allfrey},
  \citenamefont {Gota}, \citenamefont {McCarroll}, \citenamefont {Ohdachi},
  \citenamefont {Isobe}, \citenamefont {Kamio}, \citenamefont {Klumper},
  \citenamefont {Nuga}, \citenamefont {Shoji}, \citenamefont {Ziaei},
  \citenamefont {Binderbauer},\ and\ \citenamefont {Osakabe}}]{Magee2023}%
  \BibitemOpen
  \bibfield  {author} {\bibinfo {author} {\bibfnamefont {R.~M.}\ \bibnamefont
  {Magee}}, \bibinfo {author} {\bibfnamefont {K.}~\bibnamefont {Ogawa}},
  \bibinfo {author} {\bibfnamefont {T.}~\bibnamefont {Tajima}}, \bibinfo
  {author} {\bibfnamefont {I.}~\bibnamefont {Allfrey}}, \bibinfo {author}
  {\bibfnamefont {H.}~\bibnamefont {Gota}}, \bibinfo {author} {\bibfnamefont
  {P.}~\bibnamefont {McCarroll}}, \bibinfo {author} {\bibfnamefont
  {S.}~\bibnamefont {Ohdachi}}, \bibinfo {author} {\bibfnamefont
  {M.}~\bibnamefont {Isobe}}, \bibinfo {author} {\bibfnamefont
  {S.}~\bibnamefont {Kamio}}, \bibinfo {author} {\bibfnamefont
  {V.}~\bibnamefont {Klumper}}, \bibinfo {author} {\bibfnamefont
  {H.}~\bibnamefont {Nuga}}, \bibinfo {author} {\bibfnamefont {M.}~\bibnamefont
  {Shoji}}, \bibinfo {author} {\bibfnamefont {S.}~\bibnamefont {Ziaei}},
  \bibinfo {author} {\bibfnamefont {M.~W.}\ \bibnamefont {Binderbauer}},\ and\
  \bibinfo {author} {\bibfnamefont {M.}~\bibnamefont {Osakabe}},\ }\href
  {https://doi.org/10.1038/s41467-023-36655-1} {\bibfield  {journal} {\bibinfo
  {journal} {Nat. Commun.}\ }\textbf {\bibinfo {volume} {14}},\ \bibinfo
  {pages} {955} (\bibinfo {year} {2023})}\BibitemShut {NoStop}%
\bibitem [{\citenamefont {Belyaev}\ \emph {et~al.}(2005)\citenamefont
  {Belyaev}, \citenamefont {Matafonov}, \citenamefont {Vinogradov},
  \citenamefont {Krainov}, \citenamefont {Lisitsa}, \citenamefont {Roussetski},
  \citenamefont {Ignatyev},\ and\ \citenamefont {Andrianov}}]{Belyaev2005}%
  \BibitemOpen
  \bibfield  {author} {\bibinfo {author} {\bibfnamefont {V.~S.}\ \bibnamefont
  {Belyaev}}, \bibinfo {author} {\bibfnamefont {A.~P.}\ \bibnamefont
  {Matafonov}}, \bibinfo {author} {\bibfnamefont {V.~I.}\ \bibnamefont
  {Vinogradov}}, \bibinfo {author} {\bibfnamefont {V.~P.}\ \bibnamefont
  {Krainov}}, \bibinfo {author} {\bibfnamefont {V.~S.}\ \bibnamefont
  {Lisitsa}}, \bibinfo {author} {\bibfnamefont {A.~S.}\ \bibnamefont
  {Roussetski}}, \bibinfo {author} {\bibfnamefont {G.~N.}\ \bibnamefont
  {Ignatyev}},\ and\ \bibinfo {author} {\bibfnamefont {V.~P.}\ \bibnamefont
  {Andrianov}},\ }\href {https://doi.org/10.1103/PhysRevE.72.026406} {\bibfield
   {journal} {\bibinfo  {journal} {Phys. Rev. E}\ }\textbf {\bibinfo {volume}
  {72}},\ \bibinfo {pages} {026406} (\bibinfo {year} {2005})}\BibitemShut
  {NoStop}%
\bibitem [{\citenamefont {Kouhi}\ \emph {et~al.}(2011)\citenamefont {Kouhi},
  \citenamefont {Ghoranneviss}, \citenamefont {Malekynia}, \citenamefont
  {Hora}, \citenamefont {Miley}, \citenamefont {Sari}, \citenamefont {Azizi},\
  and\ \citenamefont {Razavipour}}]{Kouhi2011}%
  \BibitemOpen
  \bibfield  {author} {\bibinfo {author} {\bibfnamefont {M.}~\bibnamefont
  {Kouhi}}, \bibinfo {author} {\bibfnamefont {M.}~\bibnamefont {Ghoranneviss}},
  \bibinfo {author} {\bibfnamefont {B.}~\bibnamefont {Malekynia}}, \bibinfo
  {author} {\bibfnamefont {H.}~\bibnamefont {Hora}}, \bibinfo {author}
  {\bibfnamefont {G.}~\bibnamefont {Miley}}, \bibinfo {author} {\bibfnamefont
  {A.}~\bibnamefont {Sari}}, \bibinfo {author} {\bibfnamefont {N.}~\bibnamefont
  {Azizi}},\ and\ \bibinfo {author} {\bibfnamefont {S.}~\bibnamefont
  {Razavipour}},\ }\href {https://doi.org/10.1017/S026303461100005X} {\bibfield
   {journal} {\bibinfo  {journal} {Laser Part. Beams}\ }\textbf {\bibinfo
  {volume} {29}},\ \bibinfo {pages} {125} (\bibinfo {year} {2011})}\BibitemShut
  {NoStop}%
\bibitem [{\citenamefont {Picciotto}\ \emph {et~al.}(2014)\citenamefont
  {Picciotto}, \citenamefont {Margarone}, \citenamefont {Velyhan},
  \citenamefont {Bellutti}, \citenamefont {Krasa}, \citenamefont {Szydlowsky},
  \citenamefont {Bertuccio}, \citenamefont {Shi}, \citenamefont {Mangione},
  \citenamefont {Prokupek}, \citenamefont {Malinowska}, \citenamefont
  {Krousky}, \citenamefont {Ullschmied}, \citenamefont {Laska}, \citenamefont
  {Kucharik},\ and\ \citenamefont {Korn}}]{Picciotto2014}%
  \BibitemOpen
  \bibfield  {author} {\bibinfo {author} {\bibfnamefont {A.}~\bibnamefont
  {Picciotto}}, \bibinfo {author} {\bibfnamefont {D.}~\bibnamefont
  {Margarone}}, \bibinfo {author} {\bibfnamefont {A.}~\bibnamefont {Velyhan}},
  \bibinfo {author} {\bibfnamefont {P.}~\bibnamefont {Bellutti}}, \bibinfo
  {author} {\bibfnamefont {J.}~\bibnamefont {Krasa}}, \bibinfo {author}
  {\bibfnamefont {A.}~\bibnamefont {Szydlowsky}}, \bibinfo {author}
  {\bibfnamefont {G.}~\bibnamefont {Bertuccio}}, \bibinfo {author}
  {\bibfnamefont {Y.}~\bibnamefont {Shi}}, \bibinfo {author} {\bibfnamefont
  {A.}~\bibnamefont {Mangione}}, \bibinfo {author} {\bibfnamefont
  {J.}~\bibnamefont {Prokupek}}, \bibinfo {author} {\bibfnamefont
  {A.}~\bibnamefont {Malinowska}}, \bibinfo {author} {\bibfnamefont
  {E.}~\bibnamefont {Krousky}}, \bibinfo {author} {\bibfnamefont
  {J.}~\bibnamefont {Ullschmied}}, \bibinfo {author} {\bibfnamefont
  {L.}~\bibnamefont {Laska}}, \bibinfo {author} {\bibfnamefont
  {M.}~\bibnamefont {Kucharik}},\ and\ \bibinfo {author} {\bibfnamefont
  {G.}~\bibnamefont {Korn}},\ }\href
  {https://doi.org/10.1103/PhysRevX.4.031030} {\bibfield  {journal} {\bibinfo
  {journal} {Phys. Rev. X}\ }\textbf {\bibinfo {volume} {4}},\ \bibinfo {pages}
  {031030} (\bibinfo {year} {2014})}\BibitemShut {NoStop}%
\bibitem [{\citenamefont {Hora}\ \emph {et~al.}(2015)\citenamefont {Hora},
  \citenamefont {Korn}, \citenamefont {Giuffrida}, \citenamefont {Margarone},
  \citenamefont {Picciotto}, \citenamefont {Krasa}, \citenamefont {Jungwirth},
  \citenamefont {Ullschmied}, \citenamefont {Lalousis}, \citenamefont
  {Eliezer},\ and\ \citenamefont {et~al.}}]{Hora2015}%
  \BibitemOpen
  \bibfield  {author} {\bibinfo {author} {\bibfnamefont {H.}~\bibnamefont
  {Hora}}, \bibinfo {author} {\bibfnamefont {G.}~\bibnamefont {Korn}}, \bibinfo
  {author} {\bibfnamefont {L.}~\bibnamefont {Giuffrida}}, \bibinfo {author}
  {\bibfnamefont {D.}~\bibnamefont {Margarone}}, \bibinfo {author}
  {\bibfnamefont {A.}~\bibnamefont {Picciotto}}, \bibinfo {author}
  {\bibfnamefont {J.}~\bibnamefont {Krasa}}, \bibinfo {author} {\bibfnamefont
  {K.}~\bibnamefont {Jungwirth}}, \bibinfo {author} {\bibfnamefont
  {J.}~\bibnamefont {Ullschmied}}, \bibinfo {author} {\bibfnamefont
  {P.}~\bibnamefont {Lalousis}}, \bibinfo {author} {\bibfnamefont
  {S.}~\bibnamefont {Eliezer}},\ and\ \bibinfo {author} {\bibnamefont
  {et~al.}},\ }\href {https://doi.org/10.1017/S0263034615000634} {\bibfield
  {journal} {\bibinfo  {journal} {Laser Part. Beams}\ }\textbf {\bibinfo
  {volume} {33}},\ \bibinfo {pages} {607} (\bibinfo {year} {2015})}\BibitemShut
  {NoStop}%
\bibitem [{\citenamefont {Eliezer}\ \emph {et~al.}(2016)\citenamefont
  {Eliezer}, \citenamefont {Hora}, \citenamefont {Korn}, \citenamefont
  {Nissim},\ and\ \citenamefont {Martinez~Val}}]{Eliezer2016}%
  \BibitemOpen
  \bibfield  {author} {\bibinfo {author} {\bibfnamefont {S.}~\bibnamefont
  {Eliezer}}, \bibinfo {author} {\bibfnamefont {H.}~\bibnamefont {Hora}},
  \bibinfo {author} {\bibfnamefont {G.}~\bibnamefont {Korn}}, \bibinfo {author}
  {\bibfnamefont {N.}~\bibnamefont {Nissim}},\ and\ \bibinfo {author}
  {\bibfnamefont {J.~M.}\ \bibnamefont {Martinez~Val}},\ }\href
  {https://doi.org/10.1063/1.4950824} {\bibfield  {journal} {\bibinfo
  {journal} {Phys. Plasmas}\ }\textbf {\bibinfo {volume} {23}},\ \bibinfo
  {pages} {050704} (\bibinfo {year} {2016})}\BibitemShut {NoStop}%
\bibitem [{\citenamefont {Labaune}\ \emph {et~al.}(2016)\citenamefont
  {Labaune}, \citenamefont {Baccou}, \citenamefont {Yahia}, \citenamefont
  {Neuville},\ and\ \citenamefont {Rafelski}}]{Labaune2016}%
  \BibitemOpen
  \bibfield  {author} {\bibinfo {author} {\bibfnamefont {C.}~\bibnamefont
  {Labaune}}, \bibinfo {author} {\bibfnamefont {C.}~\bibnamefont {Baccou}},
  \bibinfo {author} {\bibfnamefont {V.}~\bibnamefont {Yahia}}, \bibinfo
  {author} {\bibfnamefont {C.}~\bibnamefont {Neuville}},\ and\ \bibinfo
  {author} {\bibfnamefont {J.}~\bibnamefont {Rafelski}},\ }\href
  {https://doi.org/10.1038/srep21202} {\bibfield  {journal} {\bibinfo
  {journal} {Sci. Rep.}\ }\textbf {\bibinfo {volume} {6}},\ \bibinfo {pages}
  {21202} (\bibinfo {year} {2016})}\BibitemShut {NoStop}%
\bibitem [{\citenamefont {Hora}\ \emph {et~al.}(2017)\citenamefont {Hora},
  \citenamefont {Eliezer}, \citenamefont {Kirchhoff}, \citenamefont {Nissim},
  \citenamefont {Wang}, \citenamefont {Lalousis}, \citenamefont {Xu},
  \citenamefont {Miley}, \citenamefont {Martinez-Val}, \citenamefont
  {McKenzie},\ and\ \citenamefont {et~al.}}]{Hora2017}%
  \BibitemOpen
  \bibfield  {author} {\bibinfo {author} {\bibfnamefont {H.}~\bibnamefont
  {Hora}}, \bibinfo {author} {\bibfnamefont {S.}~\bibnamefont {Eliezer}},
  \bibinfo {author} {\bibfnamefont {G.}~\bibnamefont {Kirchhoff}}, \bibinfo
  {author} {\bibfnamefont {N.}~\bibnamefont {Nissim}}, \bibinfo {author}
  {\bibfnamefont {J.}~\bibnamefont {Wang}}, \bibinfo {author} {\bibfnamefont
  {P.}~\bibnamefont {Lalousis}}, \bibinfo {author} {\bibfnamefont
  {Y.}~\bibnamefont {Xu}}, \bibinfo {author} {\bibfnamefont {G.}~\bibnamefont
  {Miley}}, \bibinfo {author} {\bibfnamefont {J.}~\bibnamefont {Martinez-Val}},
  \bibinfo {author} {\bibfnamefont {W.}~\bibnamefont {McKenzie}},\ and\
  \bibinfo {author} {\bibnamefont {et~al.}},\ }\href
  {https://doi.org/10.1017/S0263034617000799} {\bibfield  {journal} {\bibinfo
  {journal} {Laser Part. Beams}\ }\textbf {\bibinfo {volume} {35}},\ \bibinfo
  {pages} {730} (\bibinfo {year} {2017})}\BibitemShut {NoStop}%
\bibitem [{\citenamefont {Ruhl}\ and\ \citenamefont
  {Korn}(2022{\natexlab{a}})}]{Ruhl2022a}%
  \BibitemOpen
  \bibfield  {author} {\bibinfo {author} {\bibfnamefont {H.}~\bibnamefont
  {Ruhl}}\ and\ \bibinfo {author} {\bibfnamefont {G.}~\bibnamefont {Korn}},\
  }\href@noop {} {\bibinfo {title} {A laser-driven mixed fuel nuclear fusion
  micro-reactor concept}} (\bibinfo {year} {2022}{\natexlab{a}}),\ \Eprint
  {https://arxiv.org/abs/2202.03170} {arXiv:2202.03170 [physics.plasm-ph]}
  \BibitemShut {NoStop}%
\bibitem [{\citenamefont {Ruhl}\ and\ \citenamefont
  {Korn}(2022{\natexlab{b}})}]{Ruhl2022b}%
  \BibitemOpen
  \bibfield  {author} {\bibinfo {author} {\bibfnamefont {H.}~\bibnamefont
  {Ruhl}}\ and\ \bibinfo {author} {\bibfnamefont {G.}~\bibnamefont {Korn}},\
  }\href@noop {} {\bibinfo {title} {High current ionic flows via ultra-fast
  lasers for fusion applications}} (\bibinfo {year} {2022}{\natexlab{b}}),\
  \Eprint {https://arxiv.org/abs/2212.12941} {arXiv:2212.12941
  [physics.acc-ph]} \BibitemShut {NoStop}%
\bibitem [{\citenamefont {Mlodik}\ \emph {et~al.}(2023)\citenamefont {Mlodik},
  \citenamefont {Munirov}, \citenamefont {Rubin},\ and\ \citenamefont
  {Fisch}}]{MlodikMunirov2023}%
  \BibitemOpen
  \bibfield  {author} {\bibinfo {author} {\bibfnamefont {M.~E.}\ \bibnamefont
  {Mlodik}}, \bibinfo {author} {\bibfnamefont {V.~R.}\ \bibnamefont {Munirov}},
  \bibinfo {author} {\bibfnamefont {T.}~\bibnamefont {Rubin}},\ and\ \bibinfo
  {author} {\bibfnamefont {N.~J.}\ \bibnamefont {Fisch}},\ }\href
  {https://doi.org/10.1063/5.0140508} {\bibfield  {journal} {\bibinfo
  {journal} {Phys. Plasmas}\ }\textbf {\bibinfo {volume} {30}},\ \bibinfo
  {pages} {043301} (\bibinfo {year} {2023})}\BibitemShut {NoStop}%
\bibitem [{\citenamefont {Chen}\ \emph {et~al.}(1982)\citenamefont {Chen},
  \citenamefont {Kawai}, \citenamefont {Kawamura}, \citenamefont {Maegauchi},\
  and\ \citenamefont {Narumi}}]{Chen1982}%
  \BibitemOpen
  \bibfield  {author} {\bibinfo {author} {\bibfnamefont {W.-J.}\ \bibnamefont
  {Chen}}, \bibinfo {author} {\bibfnamefont {N.}~\bibnamefont {Kawai}},
  \bibinfo {author} {\bibfnamefont {T.}~\bibnamefont {Kawamura}}, \bibinfo
  {author} {\bibfnamefont {T.}~\bibnamefont {Maegauchi}},\ and\ \bibinfo
  {author} {\bibfnamefont {H.}~\bibnamefont {Narumi}},\ }\href
  {https://doi.org/10.1143/JPSJ.51.1620} {\bibfield  {journal} {\bibinfo
  {journal} {J. Phys. Soc. Jpn.}\ }\textbf {\bibinfo {volume} {51}},\ \bibinfo
  {pages} {1620} (\bibinfo {year} {1982})}\BibitemShut {NoStop}%
\bibitem [{\citenamefont {Nozawa}\ \emph {et~al.}(2009)\citenamefont {Nozawa},
  \citenamefont {Takahashi}, \citenamefont {Kohyama},\ and\ \citenamefont
  {Itoh}}]{Nozawa2009}%
  \BibitemOpen
  \bibfield  {author} {\bibinfo {author} {\bibfnamefont {S.}~\bibnamefont
  {Nozawa}}, \bibinfo {author} {\bibfnamefont {K.}~\bibnamefont {Takahashi}},
  \bibinfo {author} {\bibfnamefont {Y.}~\bibnamefont {Kohyama}},\ and\ \bibinfo
  {author} {\bibfnamefont {N.}~\bibnamefont {Itoh}},\ }\href
  {https://doi.org/10.1051/0004-6361/200811272} {\bibfield  {journal} {\bibinfo
   {journal} {Astron. Astrophys.}\ }\textbf {\bibinfo {volume} {499}},\
  \bibinfo {pages} {661} (\bibinfo {year} {2009})}\BibitemShut {NoStop}%
\bibitem [{\citenamefont {Jung}(1994)}]{Jung1994}%
  \BibitemOpen
  \bibfield  {author} {\bibinfo {author} {\bibfnamefont {Y.-D.}\ \bibnamefont
  {Jung}},\ }\href {https://doi.org/10.1063/1.870771} {\bibfield  {journal}
  {\bibinfo  {journal} {Phys. Plasmas}\ }\textbf {\bibinfo {volume} {1}},\
  \bibinfo {pages} {785} (\bibinfo {year} {1994})}\BibitemShut {NoStop}%
\bibitem [{\citenamefont {Elwert}(1939)}]{Elwert1939}%
  \BibitemOpen
  \bibfield  {author} {\bibinfo {author} {\bibfnamefont {G.}~\bibnamefont
  {Elwert}},\ }\href {https://doi.org/https://doi.org/10.1002/andp.19394260206}
  {\bibfield  {journal} {\bibinfo  {journal} {Ann. Phys. (Berl.)}\ }\textbf
  {\bibinfo {volume} {426}},\ \bibinfo {pages} {178} (\bibinfo {year}
  {1939})}\BibitemShut {NoStop}%
\bibitem [{\citenamefont {Maxon}\ and\ \citenamefont
  {Corman}(1967)}]{Maxon1967}%
  \BibitemOpen
  \bibfield  {author} {\bibinfo {author} {\bibfnamefont {M.~S.}\ \bibnamefont
  {Maxon}}\ and\ \bibinfo {author} {\bibfnamefont {E.~G.}\ \bibnamefont
  {Corman}},\ }\href {https://doi.org/10.1103/PhysRev.163.156} {\bibfield
  {journal} {\bibinfo  {journal} {Phys. Rev.}\ }\textbf {\bibinfo {volume}
  {163}},\ \bibinfo {pages} {156} (\bibinfo {year} {1967})}\BibitemShut
  {NoStop}%
\bibitem [{\citenamefont {Haug}(1998)}]{Haug1998}%
  \BibitemOpen
  \bibfield  {author} {\bibinfo {author} {\bibfnamefont {E.}~\bibnamefont
  {Haug}},\ }\href {https://doi.org/10.1023/A:1005098624121} {\bibfield
  {journal} {\bibinfo  {journal} {Sol. Phys.}\ }\textbf {\bibinfo {volume}
  {178}},\ \bibinfo {pages} {341} (\bibinfo {year} {1998})}\BibitemShut
  {NoStop}%
\bibitem [{\citenamefont {Ochs}\ \emph {et~al.}(2023)\citenamefont {Ochs},
  \citenamefont {Munirov},\ and\ \citenamefont {Fisch}}]{OchsMunirov2023}%
  \BibitemOpen
  \bibfield  {author} {\bibinfo {author} {\bibfnamefont {I.~E.}\ \bibnamefont
  {Ochs}}, \bibinfo {author} {\bibfnamefont {V.~R.}\ \bibnamefont {Munirov}},\
  and\ \bibinfo {author} {\bibfnamefont {N.~J.}\ \bibnamefont {Fisch}},\ }\href
  {https://doi.org/10.1063/5.0147466} {\bibfield  {journal} {\bibinfo
  {journal} {Phys. Plasmas}\ }\textbf {\bibinfo {volume} {30}},\ \bibinfo
  {pages} {052508} (\bibinfo {year} {2023})}\BibitemShut {NoStop}%
\bibitem [{\citenamefont {Post}(1987)}]{Post1987}%
  \BibitemOpen
  \bibfield  {author} {\bibinfo {author} {\bibfnamefont {R.~F.}\ \bibnamefont
  {Post}},\ }\href {https://doi.org/10.1088/0029-5515/27/10/001} {\bibfield
  {journal} {\bibinfo  {journal} {Nucl. Fusion}\ }\textbf {\bibinfo {volume}
  {27}},\ \bibinfo {pages} {1579} (\bibinfo {year} {1987})}\BibitemShut
  {NoStop}%
\bibitem [{\citenamefont {Fetterman}\ and\ \citenamefont
  {Fisch}(2011)}]{Fetterman2011}%
  \BibitemOpen
  \bibfield  {author} {\bibinfo {author} {\bibfnamefont {A.~J.}\ \bibnamefont
  {Fetterman}}\ and\ \bibinfo {author} {\bibfnamefont {N.~J.}\ \bibnamefont
  {Fisch}},\ }\href {https://doi.org/10.1063/1.3631793} {\bibfield  {journal}
  {\bibinfo  {journal} {Phys. Plasmas}\ }\textbf {\bibinfo {volume} {18}},\
  \bibinfo {pages} {094503} (\bibinfo {year} {2011})}\BibitemShut {NoStop}%
\bibitem [{\citenamefont {White}\ \emph {et~al.}(2018)\citenamefont {White},
  \citenamefont {Hassam},\ and\ \citenamefont {Brizard}}]{White2018}%
  \BibitemOpen
  \bibfield  {author} {\bibinfo {author} {\bibfnamefont {R.}~\bibnamefont
  {White}}, \bibinfo {author} {\bibfnamefont {A.}~\bibnamefont {Hassam}},\ and\
  \bibinfo {author} {\bibfnamefont {A.}~\bibnamefont {Brizard}},\ }\href
  {https://doi.org/10.1063/1.5003359} {\bibfield  {journal} {\bibinfo
  {journal} {Phys. Plasmas}\ }\textbf {\bibinfo {volume} {25}},\ \bibinfo
  {pages} {012514} (\bibinfo {year} {2018})}\BibitemShut {NoStop}%
\bibitem [{\citenamefont {Rubin}\ \emph {et~al.}(2023)\citenamefont {Rubin},
  \citenamefont {Rax},\ and\ \citenamefont {Fisch}}]{Rubin2023}%
  \BibitemOpen
  \bibfield  {author} {\bibinfo {author} {\bibfnamefont {T.}~\bibnamefont
  {Rubin}}, \bibinfo {author} {\bibfnamefont {J.~M.}\ \bibnamefont {Rax}},\
  and\ \bibinfo {author} {\bibfnamefont {N.~J.}\ \bibnamefont {Fisch}},\ }\href
  {https://doi.org/10.1063/5.0145042} {\bibfield  {journal} {\bibinfo
  {journal} {Phys. Plasmas}\ }\textbf {\bibinfo {volume} {30}},\ \bibinfo
  {pages} {052501} (\bibinfo {year} {2023})}\BibitemShut {NoStop}%
\bibitem [{\citenamefont {Beuermann}(1987)}]{Beuermann1987}%
  \BibitemOpen
  \bibfield  {author} {\bibinfo {author} {\bibfnamefont {K.}~\bibnamefont
  {Beuermann}},\ }\href {https://doi.org/10.1007/BF00668147} {\bibfield
  {journal} {\bibinfo  {journal} {Astrophys. Space Sci.}\ }\textbf {\bibinfo
  {volume} {131}},\ \bibinfo {pages} {625} (\bibinfo {year}
  {1987})}\BibitemShut {NoStop}%
\bibitem [{\citenamefont {Ciotti}\ and\ \citenamefont
  {Ostriker}(2004)}]{Ciotti2004}%
  \BibitemOpen
  \bibfield  {author} {\bibinfo {author} {\bibfnamefont {L.}~\bibnamefont
  {Ciotti}}\ and\ \bibinfo {author} {\bibfnamefont {J.~P.}\ \bibnamefont
  {Ostriker}},\ }\href {https://doi.org/10.1063/1.1718467} {\bibfield
  {journal} {\bibinfo  {journal} {AIP Conf. Proc.}\ }\textbf {\bibinfo {volume}
  {703}},\ \bibinfo {pages} {276} (\bibinfo {year} {2004})}\BibitemShut
  {NoStop}%
\end{thebibliography}

\end{document}